\documentclass[letterpaper,floatfix]{revtex4}

\usepackage{graphicx}

\begin{document}

\noindent
\large
\centerline{\bf
Coherent Quantum Control of $S_2 \leftrightarrow S_1$ Internal Conversion in Pyrazine}
\centerline{\bf
via $S_0 \to S_2/S_1$ Weak Field Excitation}

\normalsize
\begin{quote}

{\bfseries Timur Grinev \cite{address}}

{\slshape \footnotesize Department Of Chemistry, Chemical Physics Theory Group, and Center for Quantum Information and Quantum Control, University of Toronto, Toronto, Ontario M5S 3H6, Canada}

{\bfseries Moshe Shapiro}

{\slshape \footnotesize Department of Chemistry and Department of Physics, University of British Columbia, Vancouver, British Columbia \\ V6T 1Z1, Canada, and Department of Chemical Physics, Weizmann Institute of Science, Rehovot 76100, Israel}

{\bfseries Paul Brumer} 

{\slshape \footnotesize Department Of Chemistry, Chemical Physics Theory Group, and Center for Quantum Information and Quantum Control, University of Toronto, Toronto, Ontario M5S 3H6, Canada}

{\bf Abstract}

Coherent control of internal conversion (IC) between the first ($S_1$) and second ($S_2$) singlet excited electronic states in pyrazine, where the $S_2$ state is populated from the ground singlet electronic state $S_0$ by weak field excitation, is examined. Control is implemented by  shaping the laser which excites  $S_2$. Excitation and IC are considered simultaneously, using the recently introduced resonance-based control approach.
Highly successful control is achieved by optimizing both the amplitude and phase profiles of the laser spectrum. The dependence of control on the properties of resonances in $S_2$ is demonstrated.

\end{quote}
\setcounter{section}{0}

\section{Introduction}

Coherent quantum control \cite{Rice-Zhao-2000,Shapiro-Brumer-2003} has been  extensively studied for a wide variety of systems and proven to be a useful approach to controlling  properties of atomic and molecular systems.  For example, in bound systems it has been used to suppress  spontaneous emission from a manifold of states \cite{Frishman-Shapiro-2001}, and to control  radiationless transitions in collinear carbonyl sulfide OCS \cite{Collinear-OCS} and in pyrazine C$_4$H$_4$N$_2$ \cite{Christopher-Pyrazine-1,Christopher-Pyrazine-2,Christopher-Pyrazine-3}.

Christopher et al. examined \cite{Christopher-Pyrazine-1,Christopher-Pyrazine-3} radiationless transitions in pyrazine from the  $S_2$  to the $S_1$ electronic state  and controlled the process by  optimizing the superposition states belonging to $S_2$.   The problem was first studied \cite{Christopher-Pyrazine-1} using a  simplified four-mode model for the pyrazine vibrational motion \cite{Borrelli-Peluso-2003}. The  optimization technique used  showed  the possibility of performing active phase control of $S_2 \leftrightarrow S_1$ interconversion, and that this control is directly related to the presence of overlapping resonances \cite{Levine-1969,Shapiro-1972} in the $S_2$ manifold. Subsequently \cite{Christopher-Pyrazine-2,Christopher-Pyrazine-3}, the full 24-dimensional vibrational motion of pyrazine \cite{Raab-1999} was considered, and the dynamical problem  solved using an efficient L\"owdin-Feshbach QP-partitioning approach. Previous control results were fully confirmed and refined,  proving the high controllability of $S_2 \leftrightarrow S_1$ internal conversion by actively exploiting the effect of quantum interferences which was shown to rely on the presence of overlapping resonances.

In Refs. \cite{Christopher-Pyrazine-1,Christopher-Pyrazine-3} coherent control was implemented for pyrazine that was  already prepared in the excited $S_2$ state.  The  $S_0 \to S_2$ excitation process was not considered, assuming instead  that the excited states in $S_2$ were already populated. Recently, we  showed the possibility of performing effective coherent control in a simple IBr diatomic model,  where we explicitly included  the exciting laser in an approach that  simultaneously considered excitation and decay to a continuum \cite{IBr-model,Shapiro-1998}. In that case we introduced an  optimization schemes  different from the simple one used in Refs. \cite{Christopher-Pyrazine-1,Christopher-Pyrazine-3} and demonstrated  the  reliance of control  on overlapping resonances. Below we considerably generalize this study to pyrazine, explicitly introducing the  laser to excite the 24-dimensional $S_1 + S_2$ vibronic pyrazine model  \cite{Christopher-Pyrazine-2}, and using the same control and optimization schemes as for  IBr  \cite{IBr-model}. Significantly,  we confirm the dependence of controllability on the properties of the $S_2$ resonances in pyrazine.

This paper is organized as follows. Section \ref{Theory-General}  reviews the theory  explicitly accounting for the exciting laser in the weak field limit. Section \ref{Theory-Control} introduces the coherent control approach for the $S_2$ population,  points out its connection with the properties of $S_2$ resonances, and provides additional details of the approach. Section \ref{Comp-Results} provides computational results for control of pyrazine internal conversion. Section \ref{Summary} provides a summary and conclusions.

\section{$S_0 \to S_2$ Excitation and $S_2 \leftrightarrow S_1$ Internal Conversion} \label{Theory-General}

Below, $| \kappa \rangle$ denotes  vibrational states belonging to the $S_2$ electronic state, with  corresponding projection operator $Q = \sum_\kappa | \kappa \rangle \langle \kappa |$. Since the $| \kappa \rangle$ states are not eigenstates of the full Pyrazine Hamiltonian,  the system  evolves in time if it were prepared in these states.  Hence, such states  are termed resonances.  The states $| \beta \rangle$ denote  vibrational states belonging to the $S_1$ electronic state, with $P = \sum_\beta | \beta \rangle \langle \beta |$ being the associated projection operator. The full vibronic states, which are eigenstates of the full Pyrazine system,  are denoted $| \gamma \rangle$, so that $P + Q = I = \sum_\gamma | \gamma \rangle \langle \gamma |$.

\subsection{Time Evolution of the System Assumed  Already Excited} \label{Theory-General-Already-Excited}

In Refs. \cite{Christopher-Pyrazine-1,Christopher-Pyrazine-2,Christopher-Pyrazine-3}  $S_0 \to S_2$ laser excitation is assumed   to allow preparation of a  superposition of $| \kappa \rangle$ resonances:
\begin{equation}
| \Psi(0) \rangle = \sum_{\kappa'} c_{\kappa'} | \kappa' \rangle. \label{Psi_0_C_controlled}
\end{equation}
The dynamics of internal conversion was then described by an action of the propagator $U(t)$ on $| \Psi(0) \rangle$: $| \Psi(t) \rangle = U(t) | \Psi(0) \rangle$. Because $| \gamma \rangle$ are exact states of the system Hamiltonian, the spectral resolution of the evolution operator $U(t) = \exp(-i H t/\hbar)$ is $U(t) = \sum_\gamma \exp(-i E_\gamma t/\hbar) | \gamma \rangle \langle \gamma |$. This gives
\begin{equation}
| \Psi(t) \rangle = \sum_{\kappa'} c_{\kappa'} \sum_\gamma \exp(-i E_\gamma t/\hbar) \langle \gamma | \kappa' \rangle | \gamma \rangle = \sum_\gamma a_\gamma \exp(-i E_\gamma t/\hbar) | \gamma \rangle, \label{Psi_t_a_expansion}
\end{equation}
where $a_\gamma \equiv \sum_{\kappa'} c_{\kappa'} \langle \gamma | \kappa' \rangle$.

The $S_2$ electronic state population $P_{S_2}$ at time $t$ is an observable defined by the projection operator $Q$ onto the state $| \Psi(t) \rangle$:
\begin{equation}
P_{S_2} (t) = \langle \Psi (t) | Q | \Psi (t) \rangle = \sum_{\gamma',\gamma''} \tilde{a}_{\gamma'}^* (t) \, \tilde{a}_{\gamma''} (t) Q_{\gamma',\gamma''}, \label{P_S_2_t_a_scalar_expansion}
\end{equation}
where $\tilde{a}_{\gamma} (t) \equiv a_{\gamma} \exp(-i E_{\gamma} t /\hbar)$ and $Q_{\gamma',\gamma''} \equiv \langle \gamma' | Q | \gamma'' \rangle$. Equation (\ref{P_S_2_t_a_scalar_expansion}) can be rewritten in matrix form as:
\begin{equation}
P_{S_2} (t) =  \mathbf{a}^\dagger \underline{\underline{\mathbf{e}}}^{i E t/\hbar} \mathbf{Q} \, \underline{\underline{\mathbf{e}}}^{-i E t/\hbar} \mathbf{a}, \label{P_S_2_t_a_matrix_expansion-1}
\end{equation}
where  $\mathbf{a}$ is a vector with $a_\gamma$ components, $\underline{\underline{\mathbf{e}}}^{\pm i E t/\hbar}$ are square diagonal matrices composed of $\exp(\pm i E_\gamma t/\hbar)$ values, and $\mathbf{Q}$ is a square matrix with $Q_{\gamma',\gamma''}$ matrix elements.

Since $Q = \sum_\kappa | \kappa \rangle \langle \kappa |$, the matrix elements  $Q_{\gamma',\gamma''} = \langle \gamma' | Q | \gamma'' \rangle = \sum_\kappa \langle \gamma' | \kappa \rangle \langle \gamma'' | \kappa \rangle^*$. Introducing the matrix $\mathbf{R}$ with $R_{\gamma,\kappa} = \langle \gamma | \kappa \rangle$, then $Q_{\gamma',\gamma''} = \sum_\kappa R_{\gamma',\kappa} R_{\gamma'',k}^* = \sum_\kappa R_{\gamma',\kappa} R_{\kappa,\gamma''}^\dagger = [R R^\dagger]_{\gamma',\gamma''}$, giving
\begin{equation}
\mathbf{Q} = \mathbf{R} \mathbf{R}^\dagger. \label{Q_eq_R_R_dagger}
\end{equation}
In  turn, according to Eq. (\ref{Psi_t_a_expansion}), the vector $\mathbf{a}$ can be written as
\begin{equation}
\mathbf{a} = \mathbf{R} \mathbf{c}, \label{a_eq_R_c}
\end{equation}
where $\mathbf{c}$ is a vector composed of $c_{\kappa'}$ coefficients. Inserting Eqs. (\ref{Q_eq_R_R_dagger}) and (\ref{a_eq_R_c}) into Eq. (\ref{P_S_2_t_a_matrix_expansion-1}) gives:
\begin{eqnarray}
P_{S_2} (t) & = & \mathbf{c}^\dagger \mathbf{R}^\dagger \underline{\underline{\mathbf{e}}}^{i E t/\hbar} \mathbf{R} \mathbf{R}^\dagger \underline{\underline{\mathbf{e}}}^{-i E t/\hbar} \mathbf{R} \mathbf{c} \equiv \mathbf{c}^\dagger \mathbf{M}^{c \dagger} (t) \mathbf{M}^c (t) \mathbf{c} \equiv \mathbf{c}^\dagger \mathbf{K}^c (t) \mathbf{c} \nonumber \\
& = & \sum_{\kappa',\kappa''} c^*_{\kappa'} c_{\kappa''} K^c_{\kappa',\kappa''} (t) = \sum_{\kappa'} | c_{\kappa'} |^2 K^c_{\kappa',\kappa'} (t) + \sum_{\kappa' \ne \kappa''} c^*_{\kappa'} c_{\kappa''} K^c_{\kappa',\kappa''} (t), \label{P_S_2_t_a_matrix_expansion-2}
\end{eqnarray}
where $\mathbf{M}^c (t)$ and $\mathbf{K}^c (t)$ matrices are defined as
\begin{equation}
\mathbf{M}^c (t) \equiv \mathbf{R}^\dagger \underline{\underline{\mathbf{e}}}^{-i E t/\hbar} \mathbf{R}, \label{M_c_t_definition} \qquad \mathbf{K}^c (t) \equiv \mathbf{M}^{c \dagger} (t) \mathbf{M}^c (t) = \mathbf{R}^\dagger \underline{\underline{\mathbf{e}}}^{i E t/\hbar} \mathbf{R} \mathbf{R}^\dagger \underline{\underline{\mathbf{e}}}^{-i E t/\hbar} \mathbf{R}. \label{K_c_t_definition}
\end{equation}
The matrix elements of $\mathbf{M}^c(t)$ have the form
\begin{equation}
M^c_{\kappa,\kappa'} (t) = \sum_\gamma \langle \kappa | \gamma \rangle \langle \gamma | \kappa' \rangle \exp (-i E_\gamma t/\hbar) = \langle \kappa | U(t) | \kappa' \rangle, \label{M_c_t_matrix_element}
\end{equation}
being matrix elements of the $U(t)$ propagator operating between the resonances $| \kappa \rangle$ and $| \kappa' \rangle$. According to Eq. (\ref{M_c_t_matrix_element}), $M^c_{\kappa,\kappa'} (t) \neq 0$ for $\kappa \neq \kappa'$, only if there is at least one  state $| \gamma \rangle$ such that $\langle \kappa | \gamma \rangle \neq 0$ and $\langle \gamma | \kappa' \rangle \neq 0$. If so, then resonances $| \kappa \rangle$ and $| \kappa' \rangle$ are said to be overlapping. This resonance overlap property is crucial for $\mathbf{M}^c (t)$ nondiagonality which, in turn, provides $\mathbf{K}^c (t)$ nondiagonality, which allows efficient phase control of $P_{S_2} (t)$ in Eq. (\ref{P_S_2_t_a_matrix_expansion-2}) by means of phases $\varphi_{\kappa'}$ of complex coefficients $c_{\kappa'} = |c_{\kappa'}| \exp(i \varphi_{\kappa'})$ \cite{Christopher-Pyrazine-1,Christopher-Pyrazine-3}. Such phase control is termed active control, in contrast to passive control, which is control via the  $|c_{\kappa'}|$ amplitudes only.

In the case of pyrazine, which has 24 vibrational degrees of freedom, there is a large number of  $| \gamma \rangle$ states \cite{Christopher-Pyrazine-2,Raab-1999}. To make the computations feasible, instead of exact states, a set of approximate coarse-grained states is used to compute the time evolution. Specifically, the energy axis is divided into small bins $I_\alpha$, of size $\Delta_\alpha$, center energy $E_\alpha$ and density of states $\rho_\alpha$. The projector onto the coarse-grained state $| \alpha \rangle$ is then defined as:
$$| \alpha \rangle \langle \alpha | = (1/(\rho_\alpha \Delta_\alpha)) \sum_{\gamma \in I_\alpha} | \gamma \rangle \langle \gamma |,~~{\rm hence}~~ \sqrt{\rho_\alpha \Delta_\alpha} | \alpha \rangle \langle \alpha | \sqrt{\rho_\alpha \Delta_\alpha} =  \sum_{\gamma \in I_\alpha} | \gamma \rangle \langle \gamma |.$$
Thus, the coarse-grained state $| \alpha \rangle$ effectively replaces all the $| \gamma \rangle$ states in the bin $I_\alpha$. Numerically, the weighted states $ | \overline{\alpha} \rangle  \equiv \sqrt{\rho_\alpha \Delta_\alpha} | \alpha \rangle $ and their overlaps with resonances $| \kappa \rangle$ are available through our iterative solution method for pyrazine, based on QP-partitioning algorithm (described in detail in Ref. \cite{Christopher-Pyrazine-2}), giving
\begin{equation}
| \overline{\alpha} \rangle \langle \overline{\alpha} | = \sum_{\gamma \in I_\alpha} | \gamma \rangle \langle \gamma |. \label{alpha_bar_projector_definition}
\end{equation}
All the $| \gamma \rangle$ states belonging to the same bin $I_\alpha$ are treated as one effective state $| \alpha \rangle$; so that
\begin{eqnarray}
M^c_{\kappa,\kappa'}(t)& = &\sum_\gamma \langle \kappa | \gamma \rangle \langle \gamma | \kappa' \rangle \exp (-i E_\gamma t/\hbar) = \sum_\alpha \sum_{\gamma \in I_\alpha} \langle \kappa | \gamma \rangle \langle \gamma | \kappa' \rangle \exp (-i E_\gamma t/\hbar) \nonumber \\
&\approx & \sum_\alpha \langle \kappa | \alpha \rangle \langle \alpha | \kappa' \rangle \rho_\alpha \Delta_\alpha \cdot \frac{1}{\Delta_\alpha} \sum_{\gamma \in I_\alpha} \frac{1}{\rho_\alpha} \exp( -i E_\gamma t/\hbar). \label{LaserLessCoarseGraining}
\end{eqnarray}
The remaining inner sum over $\gamma \in I_\alpha$ in Eq. (\ref{LaserLessCoarseGraining}) is approximated by a corresponding integral:
\begin{eqnarray}
\frac{1}{\Delta_\alpha} \sum_{\gamma \in I_\alpha} \frac{1}{\rho_\alpha} \exp (-i E_\gamma t/\hbar) & \approx &
\frac{1}{\Delta_\alpha} \int^{E_\alpha + \Delta_\alpha/2}_{E_\alpha - \Delta_\alpha/2} dE_\gamma \exp (-i E_\gamma t/\hbar) \nonumber \\ & = &\exp (-i E_\alpha t/\hbar) \frac{\sin(\Delta_\alpha t/(2\hbar))}{\Delta_\alpha t/(2\hbar)} \equiv \tau_\alpha(t), \label{tauIntegral}
\end{eqnarray}
giving the final coarse-grained expression for $M^c_{\kappa,\kappa'}(t)$:
\begin{equation}
M^c_{\kappa,\kappa'}(t) \approx \sum_\alpha \langle \kappa | \overline{\alpha} \rangle \langle \overline{\alpha} | \kappa' \rangle \tau_\alpha (t) = \langle \kappa | \left[ \sum_\alpha \tau_\alpha (t) | \overline{\alpha} \rangle \langle \overline{\alpha} | \right] | \kappa' \rangle. \label{M_c_t_coarse_grained_definition}
\end{equation}
The quantity in the square brackets  is the coarse-grained approximation to the $U(t)$ propagator, and the sum is over all available $| \overline{\alpha} \rangle$ states. Equation (\ref{M_c_t_coarse_grained_definition}) is accurate for the evolution times which are not too large, i.e.,  when $| \tau_\alpha (t) | = |\sin (\Delta_\alpha t/(2 \hbar))/(\Delta_\alpha t/(2 \hbar))| \approx 1$, implying that $|t| \ll 2\hbar/\Delta_\alpha$. The resonance overlap phenomenon and the need for  nonzero coarse-grained off-diagonal $M^c_{\kappa,\kappa'}(t)$ discussed above  remains the same, except that the  $| \gamma \rangle$ states are replaced by $|\overline{\alpha} \rangle$ states.

\subsection{Time Evolution Due to  Laser Excitation} \label{Theory-General-Driven-By-Exciting-Laser}

Consider now the result of single photon excitation from the ground electronic state $S_0$, which produces the excited time-dependent wavepacket, as a superposition of $| \gamma \rangle$ states (here the subscript $p$ denotes pulse):
\begin{equation}
| \Psi_p (t) \rangle = \sum_\gamma b_{\gamma} (t) \exp (-i E_\gamma t/\hbar) | \gamma \rangle, \label{Wavepacket_general}
\end{equation}
where $b_{\gamma} (t)$ coefficients are, in general, time-dependent.

The $S_2$ electronic state population at time $t$ is given by:
\begin{equation}
P_{S_2} (t) = \langle \Psi_p (t) | Q | \Psi_p (t) \rangle =  \sum_{\gamma',\gamma''} \tilde{b}_{\gamma'}^* (t) \, \tilde{b}_{\gamma''} (t) Q_{\gamma',\gamma''}, \label{P_S_2_t_b_scalar_expansion}
\end{equation}
where $\tilde{b}_{\gamma} (t) \equiv b_{\gamma} (t) \exp(-i E_{\gamma} t /\hbar)$. Equation (\ref{P_S_2_t_b_scalar_expansion}) can be written in matrix form as
\begin{equation}
P_{S_2} (t) = \mathbf{b}^\dagger (t) \underline{\underline{\mathbf{e}}}^{i E t/\hbar} \mathbf{Q} \, \underline{\underline{\mathbf{e}}}^{-i E t/\hbar} \mathbf{b} (t), \label{P_S_2_t_b_matrix_expansion-1}
\end{equation}
where $\mathbf{b} (t)$ is a vector composed of $b_\gamma (t)$ components.

If the exciting laser pulse is weak, first-order time-dependent perturbation theory is applicable, and  the $b_{\gamma}(t)$ expansion coefficients in Eq. (\ref{Wavepacket_general}) can be written as
\begin{equation}
b_{\gamma}(t) = (i/\hbar) \langle \gamma | \mu | g \rangle \varepsilon_p(\omega_{\gamma,g},t), \label{b_FOTDPT}
\end{equation}
where $\mu$ is the dipole operator, $| g \rangle$ is the ground vibrational state on $S_0$, $\omega_{\gamma,g} \equiv (E_\gamma - E_g)/\hbar$, and $\varepsilon_p(\omega_{\gamma,g},t)$ is the finite-time Fourier transform of the $\varepsilon_p(t)$:
\begin{equation}
\varepsilon_p(\omega_{\gamma,g},t) \equiv \int^t_{-\infty} dt' \varepsilon_p(t') \exp (i \omega_{\gamma,g} t'). \label{epsilon_FTFT}
\end{equation}
Eq. (\ref{b_FOTDPT}) can be written in matrix-vector form as
\begin{equation}
\mathbf{b} (t) = \underline{\underline{\mu}} \; \underline{\varepsilon} (t), \label{b_FOTDPT_matrix-vector}
\end{equation}
where $\underline{\underline{\mu}}$ is a square diagonal matrix composed of $(i/\hbar) \langle \gamma | \mu | g \rangle$ values, and $\underline{\varepsilon} (t)$ is a vector composed of $\varepsilon_p(\omega_{\gamma,g},t)$ components.

Inserting Eqs. (\ref{Q_eq_R_R_dagger}) and  (\ref{b_FOTDPT_matrix-vector}) into Eq. (\ref{P_S_2_t_b_matrix_expansion-1}) gives, for the $P_{S_2} (t)$ population,
\begin{eqnarray}
& & P_{S_2} (t) = \underline{\varepsilon}^\dagger (t) \underline{\underline{\mu}}^\dagger \underline{\underline{\mathbf{e}}}^{i E t/\hbar} \mathbf{R} \mathbf{R}^\dagger \underline{\underline{\mathbf{e}}}^{-i E t/\hbar} \underline{\underline{\mu}} \; \underline{\varepsilon} (t) \equiv \underline{\varepsilon}^\dagger (t) \mathbf{M}^{\varepsilon \dagger} (t) \mathbf{M}^\varepsilon (t) \underline{\varepsilon} (t) \nonumber \equiv \underline{\varepsilon}^\dagger (t) \mathbf{K}^\varepsilon (t) \underline{\varepsilon} (t) \nonumber \\
& = & \sum_{\gamma'} | \varepsilon_p (\omega_{\gamma',g},t) |^2 K^\varepsilon_{\gamma',\gamma'} (t) + \sum_{\gamma'\ne \gamma''} \! \! \varepsilon_p^* (\omega_{\gamma',g},t) \varepsilon_p (\omega_{\gamma'',g},t) K^\varepsilon_{\gamma',\gamma''} (t), \label{P_S_2_t_b_matrix_expansion-2}
\end{eqnarray}
where $\mathbf{M}^\varepsilon (t)$ and $\mathbf{K}^\varepsilon (t)$ matrices are defined as
\begin{equation}
\mathbf{M}^\varepsilon (t) \equiv \mathbf{R}^\dagger \underline{\underline{\mathbf{e}}}^{-i E t/\hbar} \underline{\underline{\mu}}, \label{M_varepsilon_t_definition} \qquad \mathbf{K}^\varepsilon (t) \equiv \mathbf{M}^{\varepsilon \dagger} (t) \mathbf{M}^\varepsilon (t) . \label{K_varepsilon_t_definition}
\end{equation}
Since $\underline{\underline{\mu}}$ and $\underline{\underline{\mathbf{e}}}^{\pm i E t/\hbar}$ are diagonal, the only source of nondiagonality in Eqs. (\ref{P_S_2_t_b_matrix_expansion-2}) and (\ref{K_varepsilon_t_definition}) for $\mathbf{K}^\varepsilon (t)$ is $\mathbf{Q} = \mathbf{R} \mathbf{R}^\dagger$. Thus, phase control via the phases $\phi_\gamma (t)$ of complex $\varepsilon_p (\omega_{\gamma,g},t) = |\varepsilon_p (\omega_{\gamma,g},t)| \exp(i \phi_\gamma(t))$, depends solely on  properties of $\mathbf{Q}$.

A few comments are in order.  First, $\mathbf{R}$ is a rectangular matrix, with each $\kappa^{th}$ column composed of overlaps $R_{\gamma,\kappa} = \langle \gamma | \kappa \rangle$ of the  resonance $| \kappa \rangle$ with all  $| \gamma \rangle$ states. On the one hand, each resonance, being broadened in energy, has more than one nonzero $\langle \gamma | \kappa \rangle$ term in its $\kappa_{th}$ own column. On the other hand, if resonances $| \kappa \rangle$ and $| \kappa' \rangle$  overlap, then they have at least one common $| \gamma \rangle$  such that, for this $| \gamma \rangle$, both  $R_{\gamma,\kappa} \neq 0$ and $R_{\gamma,\kappa'} \neq 0$ simultaneously.

Second, all nonzero $\langle \gamma | \kappa \rangle$ components of each column in the $\mathbf{R}$ matrix that are related to one particular resonance $| \kappa \rangle$ form a square block centered along the main diagonal in the resulting $\mathbf{Q}= \mathbf{R} \mathbf{R}^\dagger$ matrix, filled by terms $Q_{\gamma',\gamma''} = \langle \gamma' | \kappa \rangle \langle \kappa | \gamma'' \rangle$. Thus, $\mathbf{Q}$ displays block-diagonal structure. Since each block dimensionality is larger than one due to resonance energy broadening, nondiagonal matrix elements in these blocks are generally nonzero, contributing to $\mathbf{K}^\varepsilon (t)$ nondiagonality, and thereby providing $P_{S_2} (t)$ phase control \textit{associated with the energy broadening of each particular resonance}. This kind of control will be discussed below. Furthermore, if resonances $| \kappa \rangle$ and $| \kappa' \rangle$  overlap, then the corresponding blocks  overlap, so that the $\mathbf{Q}$ matrix acquires a non-block-diagonal structure. In this case $Q_{\gamma',\gamma''}$ matrix elements belonging to two blocks simultaneously are a sum of terms borrowed from each block (produced by its corresponding resonance): $Q_{\gamma',\gamma''} = \langle \gamma' | \kappa \rangle \langle \kappa | \gamma'' \rangle + \langle \gamma' | \kappa' \rangle \langle \kappa' | \gamma'' \rangle$. Similarly, in the case of overlap of $N$ blocks, the sum contains $N$ terms: $Q_{\gamma',\gamma''} =  \sum_{\kappa = \kappa_1}^{\kappa_N} \langle \gamma' | \kappa \rangle \langle \kappa | \gamma'' \rangle$. As will be discussed below, \textit{the resonance overlap effect greatly increases the overall phase controllability in comparison with a pure resonance energy broadening effect}.

The nondiagonality  in this section (see above), is very different from that discussed in  Sect. \ref{Theory-General-Already-Excited}. Specifically,  in Eq. (\ref{P_S_2_t_a_matrix_expansion-2}), for the case when the system is already assumed to be excited, control is performed by means of the $c_{\kappa'}$ coeeficients, so that $\mathbf{a} = \mathbf{R} \, \mathbf{c}$, giving $\mathbf{K}^c (t) = \mathbf{R}^\dagger \underline{\underline{\mathbf{e}}}^{i E t/\hbar} \mathbf{R} \mathbf{R}^\dagger \underline{\underline{\mathbf{e}}}^{-i E t/\hbar} \mathbf{R}$ [Eq. (\ref{K_c_t_definition})].  This greatly simplifies the  $\mathbf{K}^c (t)$ nondiagonality dependence, effectively removing the resonance broadening effect and leaving only resonance overlap as the crucial effect that provides nondiagonality, \textit{i.e.}, phase control. By contrast, in this section,  $\mathbf{K}^\varepsilon (t) = \underline{\underline{\mu}}^\dagger \underline{\underline{\mathbf{e}}}^{i E t/\hbar} \mathbf{R} \mathbf{R}^\dagger \underline{\underline{\mathbf{e}}}^{-i E t/\hbar} \underline{\underline{\mu}}$ [Eq. (\ref{K_varepsilon_t_definition})] and nondiagonality is provided only by the $\mathbf{Q} = \mathbf{R} \mathbf{R}^\dagger$ matrix itself, whose nondiagonality, responsible for phase control, depends on \textit{both} resonance broadening and resonance overlap effects.

It can be noted that $\mathbf{M}^\varepsilon (t) \underline{\varepsilon} (t)$ in Eq. (\ref{P_S_2_t_b_matrix_expansion-2}) is a vector composed of components
\begin{equation}
\langle \kappa | \Psi_p (t) \rangle = \sum_\gamma \varepsilon_p (\omega_{\gamma,g},t) M^\varepsilon_{\kappa,\gamma} (t) = \sum_\gamma \varepsilon_p (\omega_{\gamma,g},t) \left[ \langle \kappa | \gamma \rangle \exp (-i E_\gamma t/\hbar) \frac{i}{\hbar} \langle \gamma | \mu | g \rangle \right]. \label{kappa_Psi_p_OverlapInitial}
\end{equation}
In the case of pyrazine, transition dipole matrix elements for the $S_0 \to S_1$ excitation are an order of magnitude smaller than for the $S_0 \to S_2$ excitation \cite{Christopher-Pyrazine-1,Christopher-Pyrazine-2,Christopher-Pyrazine-3,ChemPhysLett-2009}, thus allowing   the following ``doorway" approximation:
\begin{equation}
\langle \gamma | \mu | g \rangle = \langle \gamma | (P + Q) \mu | g \rangle = \sum_\beta \langle \gamma | \beta \rangle \langle \beta | \mu | g \rangle + \sum_\kappa \langle \gamma | \kappa \rangle \langle \kappa | \mu | g \rangle \approx \sum_\kappa \langle \gamma | \kappa \rangle \langle \kappa | \mu | g \rangle. \label{S0ToS2Transition}
\end{equation}
Equation (\ref{S0ToS2Transition}) indicates that the excitation to a full vibronic state $| \gamma \rangle$ takes place by means of a preliminary intermediate transition to a manifold of $| \kappa \rangle$ resonances. Inserting Eq. (\ref{S0ToS2Transition}) into Eq. (\ref{kappa_Psi_p_OverlapInitial}) gives
\begin{equation}
\langle \kappa | \Psi_p (t) \rangle = \sum_\gamma \varepsilon_p (\omega_{\gamma,g},t) \left[ \langle \kappa | \gamma \rangle  \exp (-i E_\gamma t/\hbar) \frac{i}{\hbar} \sum_{\kappa'} \langle \gamma | \kappa' \rangle \langle \kappa' | \mu | g \rangle \right], \label{kappa_Psi_p_OverlapExtended}
\end{equation}
which can be rewritten as
\begin{equation}
\langle \kappa | \Psi_p (t) \rangle = \sum_{\kappa'} \frac{i}{\hbar} \langle \kappa' | \mu | g \rangle \left[ \sum_\gamma \varepsilon_p(\omega_{\gamma,g},t) \langle \kappa | \gamma \rangle \langle \gamma | \kappa' \rangle \exp (-i E_\gamma t/\hbar) \right]. \label{kappa_Psi_p_OverlapIntermediate}
\end{equation}

In order to make the computations below feasible, we introduce here a coarse-graining procedure for the quantity in square brackets in Eq. (\ref{kappa_Psi_p_OverlapIntermediate}). This procedure is  similar to the one made in Ref. \cite{Christopher-Pyrazine-2}, taking into account Eqs. (\ref{LaserLessCoarseGraining}) and  (\ref{tauIntegral}). Namely, $\sum_\gamma$ is written as $\sum_\alpha \sum_{\gamma \in I_\alpha}$:
\begin{eqnarray}
\sum_\alpha \sum_{\gamma \in I_\alpha} \varepsilon_p(\omega_{\gamma,g},t) \langle \kappa | \gamma \rangle \langle \gamma | \kappa' \rangle \exp (-i E_\gamma t/\hbar) & \approx &
 \sum_\alpha \varepsilon_p (\omega_{\alpha,g},t) \langle \kappa | \alpha \rangle \langle \alpha | \kappa' \rangle \rho_\alpha \Delta_\alpha \cdot \frac{1}{\Delta_\alpha} \sum_{\gamma \in I_\alpha} \frac{1}{\rho_\alpha} \exp (-i E_\gamma t/\hbar) \nonumber \\ &\approx &
  \sum_\alpha \varepsilon_p (\omega_{\alpha,g},t) \langle \kappa | \overline{\alpha} \rangle \langle \overline{\alpha} | \kappa' \rangle \tau_\alpha(t), \label{LaserPresentCoarseGraining}
\end{eqnarray}
where $\omega_{\alpha,g} \equiv (E_\alpha - E_g)/\hbar$. Inserting Eq. (\ref{LaserPresentCoarseGraining}) into Eq. (\ref{kappa_Psi_p_OverlapIntermediate})  gives:
\begin{equation}
\langle \kappa | \Psi_p (t) \rangle \approx \sum_\alpha \varepsilon_p (\omega_{\alpha,g},t) \left[ \langle \kappa | \overline{\alpha} \rangle \, \tau_\alpha(t) \, \frac{i}{\hbar} \sum_{\kappa'} \langle \overline{\alpha} | \kappa' \rangle \langle \kappa' | \mu | g \rangle \right] \equiv \sum_\alpha \varepsilon_p (\omega_{\alpha,g},t) M^{\varepsilon, \alpha}_{\kappa,\alpha} (t). \label{kappa_Psi_p_Overlap}
\end{equation}
Below, a superscript $\alpha$ indicates the coarse-grained nature of the corresponding values. Here, the quantity
\begin{equation}
M^{\varepsilon, \alpha}_{\kappa,\alpha} (t) \equiv \langle \kappa | \overline{\alpha} \rangle \, \tau_\alpha(t) \, \frac{i}{\hbar} \sum_{\kappa'} \langle \overline{\alpha} | \kappa' \rangle \langle \kappa' | \mu | g \rangle \equiv \langle \kappa | \left[ \, \tau_\alpha(t) | \overline{\alpha} \rangle \langle \overline{\alpha} | \, \right] \left[ \sum_{\kappa'} \frac{i}{\hbar} \langle \kappa' | \mu | g \rangle | \kappa' \rangle \right] \label{M_varepsilon_kappa_alpha}
\end{equation}
is a coarse-grained version of $M^\varepsilon_{\kappa,\gamma} (t)$ [Eq. (\ref{kappa_Psi_p_OverlapInitial})] and depends only on the material system properties. If one defines
\begin{eqnarray}
\mu^\alpha_\alpha & \equiv & \frac{i}{\hbar} \sum_{\kappa'} \langle \overline{\alpha} | \kappa' \rangle \langle \kappa' | \mu | g \rangle = \frac{i}{\hbar} \langle \overline{\alpha} | \mu | g \rangle, \label{mu_alpha_def} \qquad R^\alpha_{\alpha,\kappa} \equiv \langle \overline{\alpha} | \kappa \rangle,
\end{eqnarray}
then
\begin{equation}
\mathbf{M}^{\varepsilon, \alpha} (t) = \mathbf{R}^{\alpha \dagger} \, \underline{\underline{\tau}}^\alpha (t) \, \underline{\underline{\mu}}^\alpha, \label{M_alpha_varepsilon_t_definition} \qquad
\mathbf{K}^{\varepsilon, \alpha} (t) = \mathbf{M}^{\varepsilon, \alpha \dagger} (t) \mathbf{M}^{\varepsilon, \alpha} (t) , \label{K_alpha_varepsilon_t_definition}
\end{equation}
where $\underline{\underline{\tau}}^\alpha (t)$ is a square diagonal matrix composed of $\tau_\alpha(t)$ values, and $\underline{\underline{\mu}}^\alpha$ is a square diagonal matrix composed of $(i/\hbar) \langle \overline{\alpha} | \mu | g \rangle$ values. Then the $P_{S_2} (t)$ population in terms of coarse-grained values becomes
\begin{eqnarray}
P_{S_2} (t)& = &\underline{\varepsilon}^{\alpha \dagger} (t) \mathbf{M}^{\varepsilon, \alpha \dagger} (t) \mathbf{M}^{\varepsilon, \alpha} (t) \, \underline{\varepsilon}^\alpha (t) \equiv \underline{\varepsilon}^{\alpha \dagger} (t) \mathbf{K}^{\varepsilon, \alpha} (t) \, \underline{\varepsilon}^\alpha (t) \nonumber \\
& = &  \sum_{\alpha'} | \varepsilon_p (\omega_{\alpha',g},t) |^2 K^{\varepsilon, \alpha}_{\alpha',\alpha'} (t) + \sum_{\alpha' \ne \alpha''} \! \! \varepsilon_p^* (\omega_{\alpha',g},t) \varepsilon_p (\omega_{\alpha'',g},t) K^{\varepsilon, \alpha}_{\alpha',\alpha''} (t), \label{P_S_2_t_alpha_matrix_expansion}
\end{eqnarray}
where $\underline{\varepsilon}^\alpha (t)$ is a vector composed of $\varepsilon_p (\omega_{\alpha,g},t)$ components.

The quantities $\underline{\underline{\mu}}^\alpha$ and $\underline{\underline{\tau}}^\alpha (t)$ are diagonal matrices, so the only origin of nondiagonality in Eq. (\ref{K_alpha_varepsilon_t_definition}) for $\mathbf{K}^{\varepsilon, \alpha} (t)$ and Eq. (\ref{P_S_2_t_alpha_matrix_expansion}) is via the $\mathbf{Q}^\alpha = \mathbf{R}^\alpha \mathbf{R}^{\alpha \dagger}$ matrix, composed of $Q^\alpha_{\alpha',\alpha''} = \langle \overline{\alpha}' | Q | \overline{\alpha}'' \rangle$ matrix elements. Hence,  all the $P_{S_2} (t)$ phase control considerations from above remain the same, except that $| \gamma \rangle$ states are replaced by $|\overline{\alpha} \rangle$ states. Namely, \textit{phase control is driven both by resonance energy broadening and resonance overlap. The resonance overlap effect, providing a non-block-diagional structure of} $\mathbf{Q}^\alpha$ \textit{and} $\mathbf{K}^{\varepsilon, \alpha} (t)$, \textit{strongly enhances the effect of resonance broadening}.

\section{Coherent Control of Pyrazine Internal Conversion} \label{Theory-Control}

Section \ref{Theory-General-Driven-By-Exciting-Laser} above  describes resonance broadening and resonance overlap,  two effects related to  $\mathbf{Q}$ ($\mathbf{Q}^\alpha$) and $\mathbf{K}^\varepsilon (t)$ ($\mathbf{K}^{\varepsilon, \alpha} (t)$) nondiagonality. Here, a control scheme based on resonance broadening is discussed  in Sect. \ref{Theory-Control-Single-Resonance}.  Section \ref{Theory-Control-Overlapping-Resonances} discusses a  control scheme relying on presence of resonance overlap.

\subsection{Control Associated with Single Resonance} \label{Theory-Control-Single-Resonance}

In the case of pure resonance broadening without resonance overlap,  one particular resonance $| \kappa \rangle$ has nonzero $R_{\gamma,\kappa} = \langle \gamma | \kappa \rangle$ terms for some specific set $\{ \gamma \}_{\kappa}$ of $| \gamma \rangle$ states. This results in the simplified expressions for $\mathbf{K}^\varepsilon (t)$ matrix elements for this $\{ \gamma \}_{\kappa}$ set, with the summation over $\kappa$  reduced to a single term
\begin{equation}
K^\varepsilon_{\gamma',\gamma'} (t) = |M^\varepsilon_{\kappa,\gamma'} (t)|^2, \qquad K^\varepsilon_{\gamma',\gamma''} (t) = M^{\varepsilon *}_{\kappa,\gamma'} (t) M^\varepsilon_{\kappa,\gamma''} (t) \label{K_gamma_single_resonance}
\end{equation}
for the diagonal and nondiagonal matrix elements, respectively.

The probability $P_{S_2} (t)$  [Eq. (\ref{P_S_2_t_b_matrix_expansion-2})] is a quadratic form of complex time-dependent variables $\varepsilon_p(\omega_{\gamma,g},t)$. When the pulse is already over (at $t = T_{over}$), these values become infinite-time Fourier transforms of this laser pulse at different frequencies, $\varepsilon_p(\omega_{\gamma,g})$; they are no longer time-dependent for $t \ge T_{over}$. Here we use the so-called absolute control scheme for $P_{S_2} (t)$ optimization, with the $\mathbf{K}^\varepsilon (t)$ matrix given in Eq. (\ref{K_gamma_single_resonance}). Namely, $P_{S_2} (t)$ is optimized at a desired optimization time $t = T$,
while keeping the  total energy of the pulse at $2\pi E_0$:
\begin{equation}
\sum_{\{ \gamma \}_{\kappa}} | \varepsilon_p (\omega_{\gamma,g}) |^2 = \underline{\varepsilon}^\dagger \underline{\varepsilon} = 2 \pi E_0. \label{Absolute-Control-Energy-Constraint-gamma}
\end{equation}
 This is done by  introducing  the corresponding Lagrange multiplier $\lambda^A$ (superscript $A$ denotes absolute) with the corresponding optimization function at  time $T$ defined as:
\begin{equation}
P^{\lambda; A}_{S_2} (T, \underline{\varepsilon}) = \underline{\varepsilon}^\dagger \mathbf{K}^\varepsilon (T) \underline{\varepsilon} - \lambda^A (\underline{\varepsilon}^\dagger \underline{\varepsilon} - 2 \pi E_0). \label{P_lambda_A_function}
\end{equation}
We then search for $P^{\lambda; A}_{S_2} (T, \underline{\varepsilon})$ extrema with respect to $\underline{\varepsilon}$:
\begin{equation}
\left\{
\begin{array}{l}
\displaystyle \frac{\partial P^{\lambda; A}_{S_2} (T, \underline{\varepsilon})}{\partial \, \mathrm{Re} \, [ \varepsilon_p(\omega_{\gamma,g}) ] } = 0, \nonumber \\
\displaystyle \frac{\partial P^{\lambda; A}_{S_2} (T, \underline{\varepsilon})}{\partial \, \mathrm{Im} \, [ \varepsilon_p(\omega_{\gamma,g}) ] } = 0, \qquad \gamma = 1, \ldots, N_{{\{ \gamma \}}_\kappa},
 \label{partial_P_A}
\end{array}
\right.
\end{equation}
where $N_{\{ \gamma \}_\kappa}$ is the number of $| \gamma \rangle$ states in the set ${\{ \gamma \}}_\kappa$. Conditions in Eq. (\ref{partial_P_A}), applied to Eq. (\ref{P_lambda_A_function}), lead directly to an eigenvalue problem
\begin{equation}
\mathbf{K}^\varepsilon (T) \underline{\varepsilon} = \lambda^A \underline{\varepsilon}. \label{AbsoluteEigenvalueProblem-gamma}
\end{equation}
which provides a set of eigenvalues $\lambda^A$ and  corresponding eigenvectors $\underline{\varepsilon}$ with a unit norm ($\underline{\varepsilon}^\dagger \underline{\varepsilon} = 1$). Multiplication of these $\underline{\varepsilon}$ eigenvectors by $\sqrt{2 \pi E_0}$  provides the required optimized solutions.

The $\mathbf{K}^\varepsilon (t)$ matrix is such that all but one of its $N_{\{ \gamma \}_{\kappa}}$ eigenvalues are equal exactly to 0, while its last eigenvalue is equal to the sum of its diagonal elements:
\begin{equation}
\lambda^A_n = 0, \qquad n = 1, \ldots, N_{\{ \gamma \}_{\kappa}} - 1; \qquad \lambda^A_{N_{\{ \gamma \}_{\kappa}}} = \sum_{\{ \gamma \}_{\kappa}} K^\varepsilon_{\gamma,\gamma} (t) = \sum_{\{ \gamma \}_{\kappa}} |M^\varepsilon_{\kappa,\gamma} (t)|^2. \label{Eigenvalues-Absolute-gamma}
\end{equation}
This is an analytical property of the $\mathbf{K}^\varepsilon (t)$ matrix in Eq. (\ref{K_gamma_single_resonance}), so that a numerical solution of the eigenproblem in [Eq. (\ref{AbsoluteEigenvalueProblem-gamma})] is not required. Specifically,
{\it for any time $T$, $P_{S_2} (T)$ can be set to zero, using the eigenvector corresponding to zero eigenvalue}.  In terms of  the coarse-grained $| \overline{\alpha} \rangle$ states, the  results are the same with the $\{\gamma\}_\kappa$ set replaced by $\{\overline{\alpha}\}_\kappa$.

Given the simplistic nature of this solution, numerical results are neither necessary nor are they provided below. Note, however, that this type of control is possible only if the system displays isolated resonances. This can be the case in small molecules; large molecules such as pyrazine, however, display overlapping resonances throughout the spectrum, with highly unlikely regions of isolated resonance. Such systems can be controlled via an alternate mechanism, discussed below.

\subsection{Control Associated with Overlapping Resonances} \label{Theory-Control-Overlapping-Resonances}

Here we consider a second different control scheme, termed relative control.    Namely, we optimize the \textit{ratio} of $P_{S_2} (t)$ populations at times $T_2$ and $T_1$, where $T_2 > T_1 \ge T_{over}$:
\begin{equation}
\lambda^R = \frac{P_{S_2} (T_2)}{P_{S_2} (T_1)} \to \max, \min \label{RelativeControlObjective}
\end{equation}
(where superscript $R$ denotes relative). One can optimize the value of $P_{S_2} (T_2)$, keeping the value of $P_{S_2} (T_1)$ constant \cite{Preisig} and equal to some predefined value $P_0$. Here fixed $P_{S_2}(T_1) = P_0$ assures that enhanced (or diminished) $P_{S_2}$ at the target final time $T_2$ does not simply result from a stronger (or weaker) field that simply achieves control by affecting the amount of $S_2$ excited.  To do so, we consider the optimization function
\begin{equation}
P^{\lambda; R}_{S_2} (T_2, T_1, \underline{\varepsilon}) = \underline{\varepsilon}^{\dagger} \mathbf{K}^\varepsilon (T_2) \underline{\varepsilon} - \lambda^R (\underline{\varepsilon}^{\dagger} \mathbf{K}^\varepsilon (T_1) \underline{\varepsilon} - P_0), \label{P_lambda_R_function}
\end{equation}
where $\lambda^R$ is a yet unknown Lagrange multiplier. We then find $P^{\lambda; R}_{S_2} (T_2, T_1, \underline{\varepsilon})$ extrema with respect to $\underline{\varepsilon}$ leading directly to a generalized eigenvalue problem:

\begin{equation}
\mathbf{K}^\varepsilon (T_2) \underline{\varepsilon} = \lambda^R \mathbf{K}^\varepsilon (T_1) \underline{\varepsilon}. \label{GeneralizedEigenvalueProblem}
\end{equation}
Multiplying Eq. (\ref{GeneralizedEigenvalueProblem}) by $\underline{\varepsilon}^{\dagger}$ from the left gives
\begin{equation}
\underline{\varepsilon}^{\dagger} \mathbf{K}^\varepsilon (T_2) \underline{\varepsilon} = \lambda^R \underline{\varepsilon}^{\dagger} \mathbf{K}^\varepsilon (T_1) \underline{\varepsilon} \label{P_T_2_eq_lambda_R_P_T_1}.
\end{equation}
The $\lambda^R$ is real and positive because $\underline{\varepsilon}^{\dagger} \mathbf{K}^\varepsilon (T_2) \underline{\varepsilon} = P_{S_2}(T_2)$ and $\underline{\varepsilon}^{\dagger} \mathbf{K}^\varepsilon (T_1) \underline{\varepsilon} = P_{S_2}(T_1)$ are real positive values. Dividing Eq. (\ref{P_T_2_eq_lambda_R_P_T_1}) by $\underline{\varepsilon}^{\dagger} \mathbf{K}^\varepsilon (T_1) \underline{\varepsilon}$, yields $\lambda^R = P_{S_2}(T_2)/P_{S_2}(T_1)$, i.e., $\lambda^R$  is the optimized ratio of the populations of interest [Eq. (\ref{RelativeControlObjective})].

The $\mathbf{K}^\varepsilon (t)$ matrix determinant is generally nonzero at every time $t$, so that $\mathbf{K}^\varepsilon (t)$ always has an inverse $[ \mathbf{K}^\varepsilon (t) ]^{-1}$. This allows  transformation of the generalized eigenvalue problem in Eq. (\ref{GeneralizedEigenvalueProblem}) into an ordinary eigenvalue problem. To do this, we multiply the left and right sides of Eq. (\ref{GeneralizedEigenvalueProblem}) by $[\mathbf{K}^\varepsilon (T_1)]^{-1}$ from the left:
\begin{eqnarray}
\mathbf{R}^\varepsilon (T_2, T_1) \underline{\varepsilon} & = & \lambda^R \underline{\varepsilon}, \label{RelativeControlProblem} \\
\mathbf{R}^\varepsilon (T_2, T_1) & \equiv & [\mathbf{K}^\varepsilon (T_1)]^{-1} \mathbf{K}^\varepsilon (T_2). \label{RelativeControlMatrix}
\end{eqnarray}
The solution to the eigenproblem in Eq. (\ref{RelativeControlProblem}) for times $T_2 > T_1 \ge T_{over}$ is dependent only on the properties of the material system. Moreover, this solution is the best possible in the weak field case, i.e., it is optimal \cite{Kosloff-1989}. Specifically, the maximal and minimal eigenvalues $\lambda^R$ provide the entire achievable range of $P_{S_2}(T_2)/P_{S_2}(T_1)$ for a given $T_2$ and $T_1$,  obtained using the corresponding eigenvectors $\underline{\varepsilon}$.

In terms of coarse-grained states $| \overline{\alpha} \rangle$, $\underline{\varepsilon}$ is replaced by $\underline{\varepsilon}^\alpha$, and $\mathbf{K}^\varepsilon (t)$ is replaced by $\mathbf{K}^{\varepsilon, \alpha} (t)$, giving the following coarse-grained version of the optimization problem:
\begin{eqnarray}
\mathbf{R}^{\varepsilon, \alpha} (T_2, T_1) \underline{\varepsilon}^\alpha & = & \lambda^{R, \alpha} \underline{\varepsilon}^\alpha, \label{RelativeControlProblem-alpha} \\
\mathbf{R}^{\varepsilon, \alpha} (T_2, T_1) & \equiv & [\mathbf{K}^{\varepsilon, \alpha} (T_1)]^{-1} \mathbf{K}^{\varepsilon, \alpha} (T_2). \label{RelativeControlMatrix-alpha}
\end{eqnarray}

In addressing this problem computationally, we encountered numerical instability  in Eq. (\ref{RelativeControlProblem-alpha}) if the number of $| \overline{\alpha} \rangle$ states is relatively large (150--180). Namely, the condition number of $\mathbf{K}^{\varepsilon, \alpha} (t)$  tends to become very large, resulting in an ill-conditioned matrix, preventing accurate numerical construction of $\mathbf{R}^{\varepsilon, \alpha} (T_2,T_1)$ [Eq. (\ref{RelativeControlMatrix-alpha})] and its subsequent diagonalization.
 To overcome this problem, we partitioned  the energy axis   into a limited number of $N_A$ bins in Eq. (\ref{kappa_Psi_p_Overlap}), as discussed in the Appendix, giving further broadened $|\textbf{A}\rangle$ states.

Using these further broadened $|\textbf{A}\rangle$ states allows us to reformulate the
eigenproblem in Eq. (\ref{RelativeControlProblem-alpha}) as
\begin{eqnarray}
\mathbf{R}^{\varepsilon, \mathbf{A}} (T_2, T_1) \underline{\varepsilon}^\mathbf{A} & = & \lambda^{R, \mathbf{A}} \underline{\varepsilon}^\mathbf{A}, \label{RelativeControlProblem-A} \\
\mathbf{R}^{\varepsilon, \mathbf{A}} (T_2, T_1) & \equiv & [\mathbf{K}^{\varepsilon, \mathbf{A}} (T_1)]^{-1} \mathbf{K}^{\varepsilon, \mathbf{A}} (T_2), \label{RelativeControlMatrix-A}
\end{eqnarray}
where the states $|\bar\alpha\rangle$ in Eqs. (\ref{RelativeControlProblem-alpha}) and (\ref{RelativeControlMatrix-alpha}) are replaced by the further broadened states $|\textbf{A}\rangle$, as described in the Appendix.

\subsection{Numerical Correlation between Controllability and Resonance Overlap} \label{Theory-Control-Numerical-Correlation}

In general, effects of resonance energy broadening and resonance overlap are mixed together in the structure of the $\mathbf{Q}^\mathbf{A}$ and $\mathbf{K}^{\varepsilon,\mathbf{A}} (t)$ matrices. To quantitatively estimate the  $\mathbf{K}^{\varepsilon,\mathbf{A}} (t)$ nondiagonality, providing phase control,  we utilize the Hadamard measure:
\begin{equation}
H \left( \mathbf{K}^{\varepsilon,\mathbf{A}} (t) \right) = \mathrm{det} \left( \mathbf{K}^{\varepsilon,\mathbf{A}} (t) \right) / \mathrm{det} \left( \mathrm{diag} \left( \mathbf{K}^{\varepsilon,\mathbf{A}} (t) \right) \right), \label{H_R_K_def}
\end{equation}
where det denotes a determinant, and diag is the diagonal part of a matrix. Thus, $\mathrm{det} \left( \mathrm{diag} \left( \mathbf{K}^{\varepsilon,\mathbf{A}} (t) \right) \right) = \prod^{N_A}_{A=1} K^{\varepsilon,\mathbf{A}}_{A,A} (t)$. Since $\mathbf{K}^{\varepsilon,\mathbf{A}} (t)$ is a Hermitian positive-definite matrix, both $\mathrm{det} \left( \mathbf{K}^{\varepsilon,\mathbf{A}} (t) \right)$ and $\mathrm{det} \left( \mathrm{diag} \left( \mathbf{K}^{\varepsilon,\mathbf{A}} (t) \right) \right)$ are real and positive. Furthermore, $\mathrm{det} \left( \mathbf{K}^{\varepsilon,\mathbf{A}} (t) \right) \le \mathrm{det} \left( \mathrm{diag} \left( \mathbf{K}^{\varepsilon,\mathbf{A}} (t) \right) \right)$, giving
\begin{equation}
0 < H \left( \mathbf{K}^{\varepsilon,\mathbf{A}} (t) \right) \le 1, \label{H_R_margins}
\end{equation}
where the equality applies  if and only if $\mathbf{K}^{\varepsilon,\mathbf{A}} (t)$ is strictly diagonal.

The determinant of $\mathbf{R}^{\varepsilon,\mathbf{A}} (T_2,T_1)$ can be expressed as
\begin{equation}
\mathrm{det} \left( \mathbf{R}^{\varepsilon,\mathbf{A}} (T_2,T_1) \right) = \mathrm{det}\left[ \left[\mathbf{K}^{\varepsilon,\mathbf{A}} (T_1) \right]^{-1} \mathbf{K}^{\varepsilon,\mathbf{A}} (T_2) \right] = \mathrm{det} \left( \mathbf{K}^{\varepsilon,\mathbf{A}} (T_2) \right) / \mathrm{det} \left( \mathbf{K}^{\varepsilon,\mathbf{A}} (T_1) \right). \label{R_det_expression}
\end{equation}
Hadamard-like measures of non-diagonality for $\mathbf{R}^{\varepsilon,\mathbf{A}} (t)$ are introduced in a similar manner:
\begin{eqnarray}
H_R \left( \mathbf{R}^{\varepsilon,\mathbf{A}} (T_2,T_1) \right) & = & \frac{\mathrm{det} \left( \mathbf{R}^{\varepsilon,\mathbf{A}} (T_2,T_1) \right)}{\mathrm{det} \left[ \mathrm{diag} ( [ \mathbf{K}^{\varepsilon,\mathbf{A}} (T_1) \right]^{-1} ) \; \mathrm{diag}(\mathbf{K}^{\varepsilon,\mathbf{A}} (T_2)) ]} = \frac{H \left( \mathbf{K}^{\varepsilon,\mathbf{A}} (T_2) \right)}{\mathrm{det}[\mathrm{diag}([\mathbf{K}^{\varepsilon,\mathbf{A}} (T_1)]^{-1})] \cdot \mathrm{det}(\mathbf{K}^{\varepsilon,\mathbf{A}} (T_1))}, \label{H_R_R_def} \\
H_C \left( \mathbf{R}^{\varepsilon,\mathbf{A}} (T_2,T_1) \right) & = & \frac{\mathrm{det}(\mathbf{R}^{\varepsilon,\mathbf{A}} (T_2,T_1))}{\mathrm{det}(\mathrm{diag}(\mathbf{R}^{\varepsilon,\mathbf{A}} (T_2,T_1)))} = \frac{\mathrm{det}(\mathbf{K}^{\varepsilon,\mathbf{A}} (T_2))}{\mathrm{det}(\mathrm{diag}(\mathbf{R}^{\varepsilon,\mathbf{A}} (T_2,T_1))) \cdot \mathrm{det}(\mathbf{K}^{\varepsilon,\mathbf{A}} (T_1))}, \label{H_C_R_def}
\end{eqnarray}
where Eq. (\ref{R_det_expression}) is used. The  subscript $R$  denotes real, and subscript $C$  denotes complex. $H_R \left( \mathbf{R}^{\varepsilon,\mathbf{A}} (T_2,T_1) \right)$ is real because both its numerator and denominator are real.

In order to quantitatively estimate the extent of resonance overlap, we use the same overlap matrix as in Ref. \cite{Christopher-Pyrazine-3}, but include only the $| \overline{\alpha} \rangle$ states, which are populated by the exciting laser spanning the energy range $[ E_L, E_H ]$:
\begin{equation}
\Omega^\alpha_{\kappa,\kappa'} = \sum_{\alpha, \\ E_\alpha \in [ E_L, E_H ]} \left| \langle \kappa | \overline{\alpha} \rangle \right| \cdot \left| \langle \overline{\alpha} | \kappa' \rangle \right|.\end{equation}
The Hadamard non-diagonality measure for the $\mathbf{\Omega}^\alpha$ matrix of size $N_Q \times N_Q$, composed of $\Omega^\alpha_{\kappa,\kappa'}$ values, is introduced as
\begin{equation}
H \left( \mathbf{\Omega}^\alpha \right) = \mathrm{det} \left( \mathbf{\Omega}^\alpha \right) / \mathrm{det} \left( \mathrm{diag} \left( \mathbf{\Omega}^\alpha \right) \right). \label{H_Omega_def}
\end{equation}
The numerator in Eq. (\ref{H_Omega_def}) is shown numerically to be always real and positive, and the denominator  is equal to $\prod^{N_Q}_{\kappa=1} \Omega^\alpha_{\kappa,\kappa}$, and thus also real and positive.  The same inequality as in Eq. (\ref{H_R_margins}) is valid for $H \left( \mathbf{\Omega}^\alpha \right)$.

\subsection{Implementation of the Shaped Laser as a Linear Combination of Gaussian Laser Pulses} \label{Theory-Control-Gaussian-Implementation}

The eigenvector $\underline{\varepsilon}^\mathbf{A}$ providing  the desired optimized value $\lambda^{R, \mathbf{A}}$ after the pulse is over [Eq. (\ref{RelativeControlProblem-A})] is a finite discrete set of complex values of laser amplitudes $\varepsilon_p (\omega_{A,g})$, at different frequencies. These values can be reached in multiple ways. The approach  used for the IBr model  \cite{IBr-model}, is also used here: namely, to obtain the desired set of $\varepsilon_p(\omega_{A,g})$ values, $A = 1, \ldots, N_A$, it is sufficient to take the same number of linearly independent functions $\varepsilon_a(\omega)$, and expand the components of  $\underline{\varepsilon}^\mathbf{A}$  in terms of $\varepsilon_a(\omega)$ at all $\omega_{A,g}$ frequencies with the (as yet unknown) time-independent complex coefficients $d_a$:
\begin{equation}
\varepsilon_p(\omega_{A,g}) = \sum_a d_a \varepsilon_a(\omega_{A,g}), \qquad a, A = 1, \ldots, N_A, \qquad t \ge T_{over}. \label{d_inf_time_expansion}
\end{equation}
or, as a matrix equation:
\begin{equation}
\underline{\varepsilon}^\mathbf{A} = \mathbf{B} \, \mathbf{d}, \qquad B_{A,a} = \varepsilon_a(\omega_{A,g}), \qquad \mathbf{d} = (d_1, \ldots, d_{N_A})^T, \qquad t \ge T_{over}. \label{epsilon_p_eq_B_d_TI}
\end{equation}
The set of $\varepsilon_a(\omega)$ functions is linearly independent,  the  $\mathbf{B}$ determinant is nonzero, and the unique nonzero vector $\mathbf{d}$ exists as a solution of Eq. (\ref{epsilon_p_eq_B_d_TI}), found as
\begin{equation}
\mathbf{d} = [\mathbf{B}]^{-1} \underline{\varepsilon}^\mathbf{A}. \label{d_solution}
\end{equation}
The basis functions $\varepsilon_a(\omega)$ in frequency domain can be assumed to be \textit{infinite-time} Fourier transforms of the corresponding basis functions $\varepsilon_a(t)$ in time domain (the latter are all vanishing when $t \ge T_{over}$). In turn, \textit{finite-time} Fourier transforms of $\varepsilon_a(t)$ can be written as $\varepsilon_a(\omega,t)$, and at finite times Eq. (\ref{d_inf_time_expansion}) takes the form:
\begin{equation}
\varepsilon_p(\omega_{A,g},t) = \sum_a d_a \varepsilon_a(\omega_{A,g},t), \qquad a, A = 1, \ldots, N_A, \label{FiniteTmeFourierTransformExpansion}
\end{equation}
i.e.,
\begin{equation}
\underline{\varepsilon}^\mathbf{A} (t) = \mathbf{B}(t) \mathbf{d}, \qquad B_{A,a} (t) = \varepsilon_a(\omega_{A,g},t), \qquad \mathbf{d} = (d_1, \ldots, d_{N_A})^T. \label{epsilon_p_eq_B_d_TD}
\end{equation}

Using Eq. (\ref{epsilon_p_eq_B_d_TD}), $P_{S_2}(t)$, Eq. (\ref{P_S_2_t_A_matrix_expansion}), can be expressed in terms of the $\mathbf{d}$ vector:
\begin{equation}
P_{S_2} (t) = \underline{\varepsilon}^{\mathbf{A} \dagger} (t) \mathbf{K}^{\varepsilon,\mathbf{A}} (t) \underline{\varepsilon}^\mathbf{A} (t) = \mathbf{d}^\dagger \mathbf{B}^\dagger(t) \mathbf{K}^{\varepsilon,\mathbf{A}} (t) \mathbf{B}(t) \mathbf{d} . \label{P_S_2_through_d}
\end{equation}

Thus, the $\mathbf{d}$ vector in Eq. (\ref{d_solution}) can be used for time propagation of $P_{S_2} (t)$ [Eq. (\ref{P_S_2_through_d})] at all times: before the laser is turned on, while the laser is on, and after the laser is off. Optimized populations  always satisfy the condition $P_{S_2}(t = T_2) = \lambda^{R, \mathbf{A}} P_{S_2}(t = T_1)$.

To perform numerical computations, we select  a set of Gaussian laser pulses  $\varepsilon_a(t)$,  centered at different frequencies $\omega_a$:
\begin{equation}
\varepsilon_a (t) = \epsilon_a/(2 \sqrt{\pi} \alpha_a) \exp \left( -\left( t/(2\alpha_a) \right)^2 -i \omega_a t \right). \label{Gaussian_varepsilon_a_t}
\end{equation}
  The finite-time Fourier transform of this Gaussian pulse, $\varepsilon_a (\omega,t)$, [Eq. (\ref{epsilon_FTFT})], can be expressed analytically \cite{Shapiro-1993,Shapiro-Femtosecond,Wolfram-Erf} as:
\begin{eqnarray}
\varepsilon_a (\omega,t) & = & (\epsilon_a/2) \exp \left( -\alpha_a^2 \left( \omega - \omega_a \right)^2 \right) \! \left\{ 2 \! - \! \exp \! \left[ \left( \alpha_a (\omega - \omega_a) + i t/(2\alpha_a) \right)^2 \right] \! W \! \left( \alpha_a (\omega - \omega_a) + i t/(2\alpha_a) \right) \right\},
\label{epsilon_omega_t_Gaussian}
\end{eqnarray}
where $W(z)$ is the complex error function \cite{Wolfram-Erf,Abramowitz-Stegun}.  At times  $t > T_{over} =  4 \sqrt{2 \ln 2} \, \alpha_a$ this becomes
\begin{equation}
\varepsilon_a (\omega) = \epsilon_a \exp \left( -\alpha_a^2 \left( \omega - \omega_a \right)^2 \right).
\end{equation}

Using Eq. (\ref{Gaussian_varepsilon_a_t}), the control
pulse $\varepsilon_p(t)$ in time domain is
\begin{equation}
\varepsilon_p (t) = \sum_{a = 1}^{N_A} d_a \varepsilon_a (t) = \sum_{a=1}^{N_A} d_a \epsilon_a/(2 \sqrt{\pi} \alpha_a) \exp \left( -\left( t/(2\alpha_a) \right)^2 -i \omega_a t \right), \label{epsilon_p_t}
\end{equation}
with  infinite-time Fourier transform
\begin{equation}
\varepsilon_p(\omega) = \sum_{a = 1}^{N_A} d_a \varepsilon_a (\omega) = \sum_{a = 1}^{N_A} d_a \epsilon_a \exp \left( -\alpha_a^2 \left( \omega - \omega_a \right)^2 \right). \label{ControlledPulseSpectrum}
\end{equation}
By construction, the $\varepsilon_p(\omega_{A,g})$ value should be constant inside the corresponding $I_A$ bin [Eq. (\ref{kappa_Psi_p_BinnedOverlap})].  The  $\varepsilon_p(\omega)$ function [Eq. (\ref{ControlledPulseSpectrum})] is smooth and does not satisfy this requirement exactly. Nevertheless, if $N_A$ is large enough each $I_A$ bin becomes relatively small, and the smooth function in Eq. (\ref{ControlledPulseSpectrum}) in each bin can be approximately treated as constant.

\section{Computational Results} \label{Comp-Results}

Consider $S_0 \to S_2$ excitation to coherently control $S_2 \leftrightarrow S_1$ interconversion dynamics  of pyrazine excited using weak light in the perturbative regime. We use the pyrazine vibronic structure of Refs. \cite{Christopher-Pyrazine-2} and \cite{Christopher-Pyrazine-3}, and partition the energy  into 2000 bins,  in the range 4.06--6.06 eV,  where energy is referred to the ground vibrational $S_0$ state. Here, 4.06 eV is the ${S_1}$ energy at the $S_0$ nuclear equilibrium configuration \cite{Raab-1999,Batista-2006}. The $Q$ space consists of the 176 brightest (most optically accessible) $| \kappa \rangle$ resonances, having the largest values of  $\langle \kappa | \mu | g \rangle$. In this case the QP-partitioning approach gives 76775 coarse-grained vibronic states $ | \overline{\alpha} \rangle$, with energies ranging from  4.06 to 6.06 eV. Thus, there are 76775$\times$176 = 13512400 $R^\alpha_{\alpha,\kappa} = \langle \overline{\alpha} | \kappa \rangle$ values. These are used together with 176 $\langle \kappa | \mu | g \rangle$ values to compute the dynamics of interest.

\subsection{Uncontrolled Excitation and Decay Dynamics}

\begin{figure}[htp]
\begin{center}
\includegraphics[height = 11cm, width = 12cm]{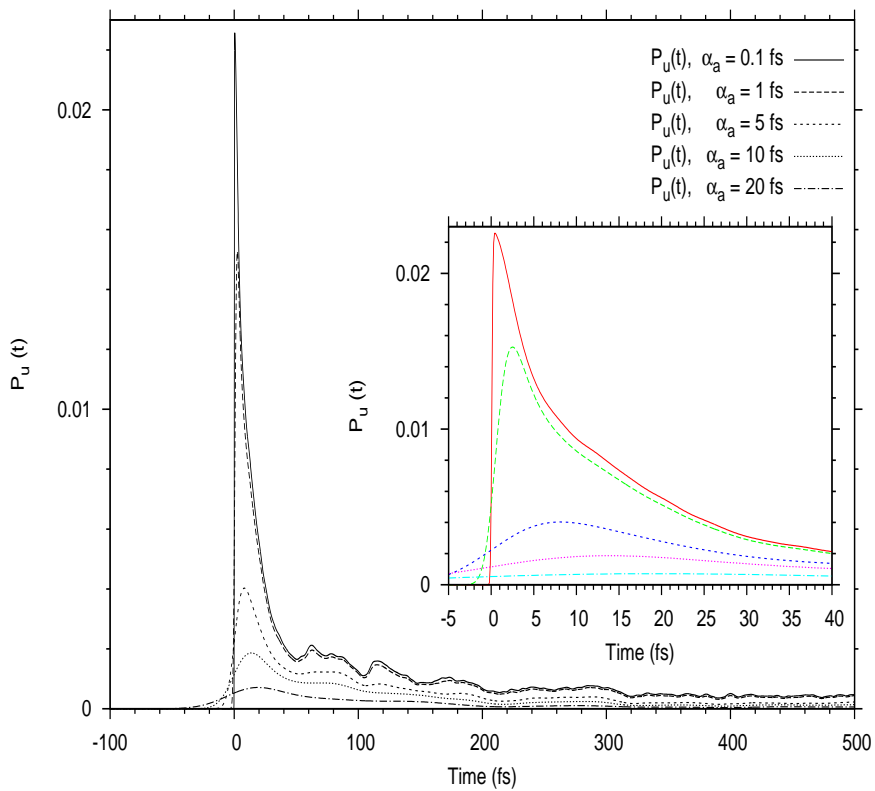}
\caption{ $S_2$ populations $P_{S_2} (t)$, denoted $P_u (t)$ here, produced by Gaussian laser pulses of different time duration. Panel inset: The same data, shown on a shorter time scale.}\label{Figure-P_u_t_Diff_alpha_a}
\end{center}
\end{figure}

Figure \ref{Figure-P_u_t_Diff_alpha_a} shows characteristic examples of $P_{S_2}(t)$ populations produced by a single Gaussian laser pulses of differing time durations,  where the subscript $u$ denotes ``uncontrolled".   These  examples are computed with the laser center frequency corresponding to 4.84 eV. It is notable that the uppermost population curve in Fig. \ref{Figure-P_u_t_Diff_alpha_a}, produced by the pulse with a time duration $\sim$1 fs ($\alpha_a = 0.1$ fs) is, at times $t > 0.5$ fs,  similar in shape to the zero-zero curve in Fig. 5,  Ref. \cite{Christopher-Pyrazine-3}. This is the case because the  ultrafast laser pulse behaves like $\epsilon_a \delta(t)$ on the femtosecond timescale,  and its finite-time Fourier transform is nearly constant, $\approx \epsilon_a$. As a consequence, in this specific case, after the pulse is over, $P_{S_2}(t)$ in Eq. (\ref{P_S_2_t_alpha_matrix_expansion}) is the same up to a constant scaling factor  as the zero-zero $P_{S_2}(t)$ in Eq. (\ref{P_S_2_t_a_matrix_expansion-2}), with $c_{\kappa'} \propto  (i/\hbar) \langle \kappa' | \mu | g \rangle \epsilon_a$.

\begin{figure}[htp]
\begin{center}
\includegraphics[height = 11cm, width = 12cm]{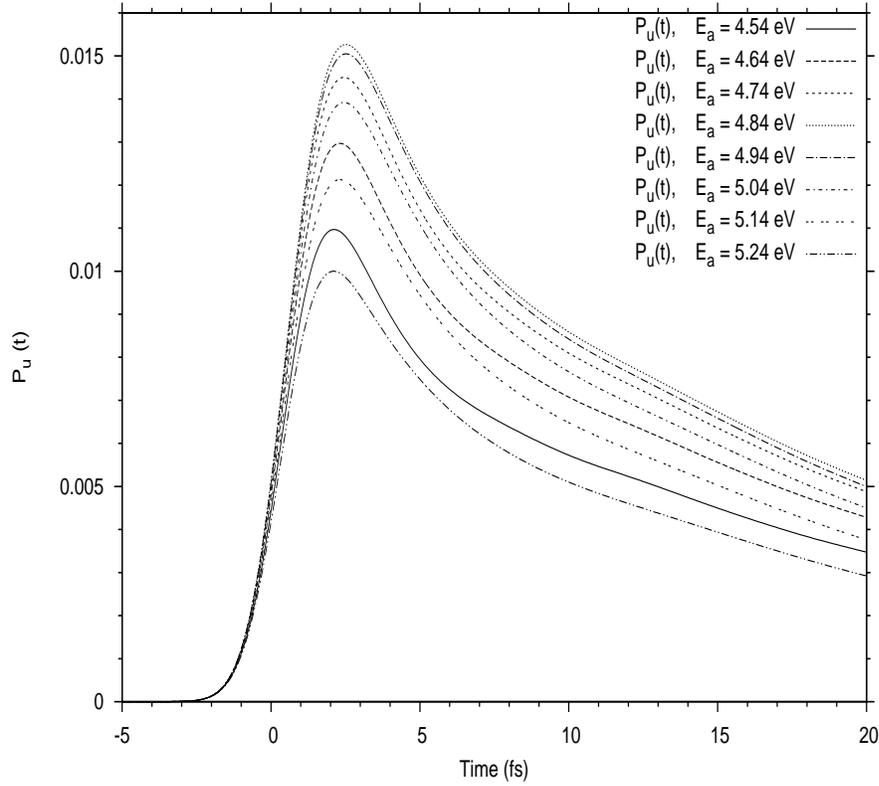}
\caption{ $S_2$ populations, $P_{S_2} (t)$, denoted $P_u (t)$ here, produced by short Gaussian laser pulses with the same $\alpha_a = 1.0$ fs, but different center frequencies.} \label{Figure-P_u_t_Short_Diff_omega_a}
\end{center}
\end{figure}

Figure \ref{Figure-P_u_t_Short_Diff_omega_a} shows $P_{S_2}(t)$ populations produced by Gaussian lasers having the same short time duration $\approx$10 fs ($\alpha_a = 1.0$ fs), but different center frequencies. In this case all  populations behave similarly on a short time scale, differing by the overall magnitude due to the difference in $\langle \kappa' | \mu | g \rangle$ values for different resonances $| \kappa' \rangle$.

\begin{figure}[htp]
\begin{center}
\includegraphics[height = 11cm, width = 12cm]{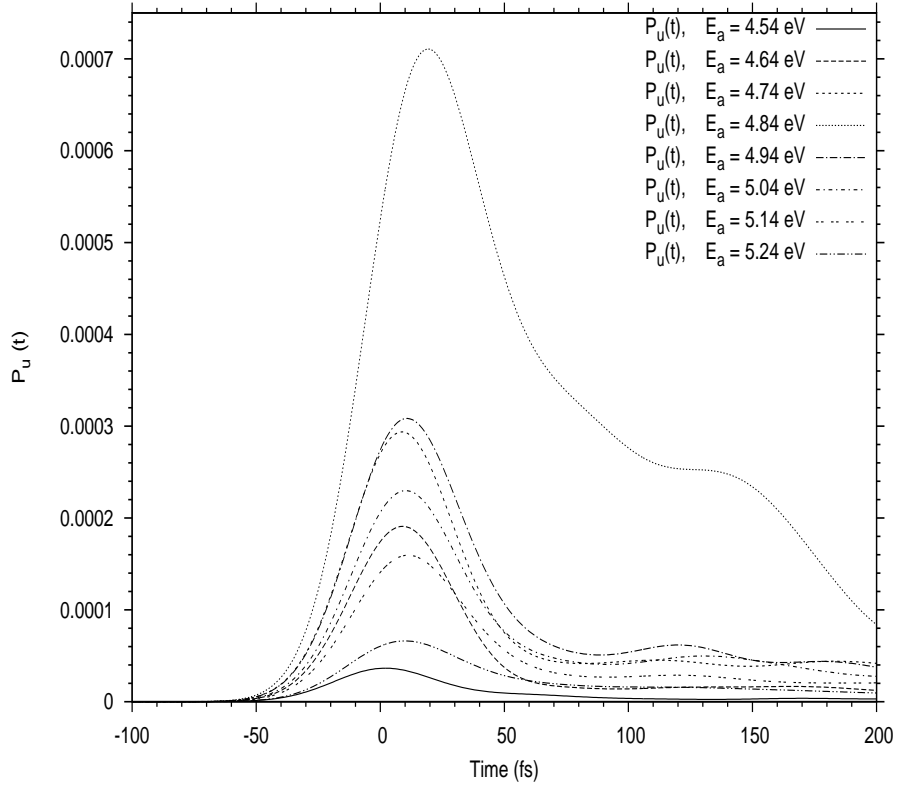}
\caption{Bottom:  $S_2$ populations, $P_{S_2} (t)$, denoted $P_u (t)$ here, produced by long Gaussian laser pulses with the same $\alpha_a = 20.0$ fs, but different center frequencies.} \label{Figure-P_u_t_Long_Diff_omega_a}
\end{center}
\end{figure}

Figure \ref{Figure-P_u_t_Long_Diff_omega_a} shows $P_{S_2}(t)$ populations produced by Gaussian lasers with long time duration around 200 fs ($\alpha_a = 20.0$ fs), using different frequencies. In contrast with Fig. \ref{Figure-P_u_t_Short_Diff_omega_a},  there are significant differences in $S_2 \leftrightarrow S_1$ IC dynamics, depending on the frequency used. Figure \ref{Figure-P_u_t_Long_Diff_omega_a} shows that the laser with 4.84 eV  photon energy produces a larger population, which also tends to decay slower, than in other cases, thus, marking the region of relative stability in pyrazine resonance structure.

Both Figs. \ref{Figure-P_u_t_Short_Diff_omega_a} and  \ref{Figure-P_u_t_Long_Diff_omega_a} qualitatively correlate well with the corresponding results for $S_0 \to S_2 \leftrightarrow S_1$ dynamics in Ref. \cite{Pyrazine-Ioannis-2}, obtained using a more general non-perturbative time-dependent dynamical approach \cite{Pyrazine-Ioannis-1}.

\subsection{Control Involving Multiple Overlapping Resonances}

Consider first  sample numerical results for  $H \left( \mathbf{\Omega}^\alpha \right)$, the measure of the extent of  resonance overlap [Eq. (\ref{H_Omega_def})] and the quantities associated with it. These quantities are $H \left( \mathbf{K}^{\varepsilon,\mathbf{A}} (t) \right)$ [Eq. (\ref{H_R_K_def})], which is the $\mathbf{K}^{\varepsilon,\mathbf{A}} (t)$ non-diagonality measure, shown at  $T_1 = 150$ fs and $T_2 = 250$ fs; and two measures of the non-diagonality  $\mathbf{R}^{\varepsilon,\mathbf{A}} (T_2,T_1)$ , $H_R \left( \mathbf{R}^{\varepsilon,\mathbf{A}} (T_2,T_1) \right)$  [Eq. (\ref{H_R_R_def})], and $|H_C \left( \mathbf{R}^{\varepsilon,\mathbf{A}} (T_2,T_1) \right)|\,$ \,[Eq. (\ref{H_C_R_def})].  In addition, we tabulate    $\lambda^{R, \mathbf{A}}_{\min}$ and $\lambda^{R, \mathbf{A}}_{\max}$, denoting minimal and maximal eigenvalues of the eigenproblem in Eq. (\ref{RelativeControlProblem-A}) and which we term ``control extents". Values for  128 $I_A$ bins (degrees of freedom of the laser), are listed in Table \ref{Table-H_Omega}.  Note first the enormous range of control possible for the ratio $P_{S_2}(T_2)/P_{S_2}(T_1)$ as indicated by the $\lambda^{R, \mathbf{A}}_{\min}$ and $\lambda^{R, \mathbf{A}}_{\max}$. For example, for the first energy interval, this ratio can range from $3.05 \times 10^{-6}$ to $3.90 \times 10^{+5}$, a range of over $1 \times 10^{+11}$.

\begin{table}[htp]
\begin{center}
\caption{The (1/128) power of $H \left( \mathbf{\Omega}^\alpha \right)$, $H \left(
\mathbf{K}^{\varepsilon,\mathbf{A}} (T_1) \right)$, $H \left(
\mathbf{K}^{\varepsilon,\mathbf{A}} (T_2) \right)$, $H_R \left(
\mathbf{R}^{\varepsilon,\mathbf{A}} (T_2,T_1) \right)$, $|H_C \left(
\mathbf{R}^{\varepsilon,\mathbf{A}} (T_2,T_1) \right)|$, as well as $\lambda^{R,
\mathbf{A}}_{\min}$ and $\lambda^{R, \mathbf{A}}_{\max}$ for different
energy intervals $[E_L, E_H]$.   Here, $T_1 = 150$ fs, $T_2 = 250$ fs.}
\begin{tabular}{llllll}
\hline \hline
$[ E_L, E_H ]$, eV
         & $[ 4.46, 4.66 ]$      & $[ 4.66, 4.86 ]$      & $[ 4.86, 5.06 ]$
     & $[ 5.06, 5.26 ]$      & $[ 5.26, 5.46 ]$      \\
\hline $H \left( \mathbf{\Omega}^\alpha \right)$
         & 1.76$\times$10$^{-1}$ & 2.77$\times$10$^{-1}$ &
3.09$\times$10$^{-1}$ & 3.15$\times$10$^{-1}$ & 2.80$\times$10$^{-1}$ \\
$H \left( \mathbf{K}^{\varepsilon,\mathbf{A}} (T_1) \right)$
         & 1.08$\times$10$^{-2}$ & 2.68$\times$10$^{-2}$ &
9.05$\times$10$^{-2}$ & 1.36$\times$10$^{-1}$ & 1.13$\times$10$^{-1}$ \\
$H \left( \mathbf{K}^{\varepsilon,\mathbf{A}} (T_2) \right)$
         & 1.23$\times$10$^{-2}$ & 2.50$\times$10$^{-2}$ &
9.45$\times$10$^{-2}$ & 1.29$\times$10$^{-1}$ & 1.06$\times$10$^{-1}$ \\
$H_R \left( \mathbf{R}^{\varepsilon,\mathbf{A}} (T_2,T_1) \right)$
         & 1.41$\times$10$^{-4}$ & 8.00$\times$10$^{-4}$ &
1.20$\times$10$^{-2}$ & 2.38$\times$10$^{-2}$ & 1.51$\times$10$^{-2}$ \\
$ \left| H_C \left( \mathbf{R}^{\varepsilon,\mathbf{A}} (T_2,T_1) \right)
\right| $ & 1.36$\times$10$^{-4}$ & 9.05$\times$10$^{-4}$ &
1.75$\times$10$^{-2}$ & 3.41$\times$10$^{-2}$ & 1.79$\times$10$^{-2}$ \\
$\lambda^{R, \mathbf{A}}_{\min}$
         & 3.05$\times$10$^{-6}$ & 3.36$\times$10$^{-5}$ &
5.54$\times$10$^{-4}$ & 1.29$\times$10$^{-3}$ & 7.30$\times$10$^{-4}$ \\
$\lambda^{R, \mathbf{A}}_{\max}$
         & 3.90$\times$10$^{+5}$ & 4.32$\times$10$^{+4}$ &
1.89$\times$10$^{+3}$ & 6.67$\times$10$^{+2}$ & 1.92$\times$10$^{+3}$ \\
\hline \hline
\end{tabular}
\label{Table-H_Omega}
\end{center}
\end{table}

The  measures in  Table I  are obtained  using  products of 128 matrix elements of the corresponding matrices.  Since each of these values is small, we report the 1/128 power of these measures.  From Table \ref{Table-H_Omega} one can see  a well defined correlation  between $H \left( \mathbf{\Omega}^\alpha \right)$ and the other quantities. Generally, when $H \left( \mathbf{\Omega}^\alpha \right)$ is small, so too are  $H \left( \mathbf{K}^{\varepsilon,\mathbf{A}} (T_1) \right)$, $H \left( \mathbf{K}^{\varepsilon,\mathbf{A}} (T_2) \right)$, $H_R \left( \mathbf{R}^{\varepsilon,\mathbf{A}} (T_2,T_1) \right)$ and $| H_C \left( \mathbf{R}^{\varepsilon,\mathbf{A}} (T_2,T_1) \right) |\,$\, (meaning a larger extent of non-diagonality in the corresponding matrices).  In particular, correlation is good with  $\lambda^{R, \mathbf{A}}_{\max} - \lambda^{R, \mathbf{A}}_{\min}$;  when it is
large, a greater extent of  coherent control is possible,  in agreement with the non-diagonality measures.

Numerically implementing controlled $P_{S_2}(t)$ dynamics  proceeded as follows. First, the eigenvalue problem in Eq. (\ref{RelativeControlProblem-A}) is numerically solved for the particular number of bins $N_A$ in the desired energy range $[ E_L, E_H]$, providing the set of eigenvalues $\lambda^{R, \mathbf{A}}$ and corresponding eigenvectors $\underline{\varepsilon}^\mathbf{A}$, which give the $\lambda^{R, \mathbf{A}}$ as $P_{S_2}(T_2)/P_{S_2}(T_1)$ ratios during the $P_{S_2}(t)$ time propagation ($T_2 > T_1 \ge T_{over}$). Then, a set of linearly independent $N_A$ Gaussian lasers [Eqs. (\ref{Gaussian_varepsilon_a_t})--(\ref{ControlledPulseSpectrum})], is introduced (all with the same $\alpha_a$), contiguously and uniformly covering the desired energy range $[E_L,E_H]$.  The  eigenvectors  obtained $\underline{\varepsilon}^\mathbf{A}$ are then
expanded in terms of this Gaussian basis with the $\mathbf{d}$ coefficients given by  Eq. (\ref{d_solution}).  The dynamics are then propagated from $t \le -T_{over}$ to $t \ge T_2$ using the corresponding $\mathbf{d}$ coefficients for each $\underline{\varepsilon}^\mathbf{A}$ eigenvector; finite-time Fourier transforms of the pulses in Eq. (\ref{FiniteTmeFourierTransformExpansion}) are produced using Eq. (\ref{epsilon_omega_t_Gaussian}), and the pulse time profiles are given in Eq. (\ref{epsilon_p_t}). The perturbative nature of the dynamics makes it possible to scale $P_{S_2}(t)$ uniformly by  multiplying the $\underline{\varepsilon}^\mathbf{A}$ eigenvector by a scalar constant. We utilized this scaling option to allow presentation of both the maximization and minimization results to be shown on the same figure (upper panel, Fig. \ref{Figure-P_c_t_128_bins}) below.  Specifically, the maximization curve is multiplied throughout by $8.3 \times 10^{-5}$

 An experimental suggestion of R. J. Gordon  (University of Illinois, Chicago) prompted our using a controllable laser in the wavelength range 250--265 nm, with  time duration $\sim$150--200 fs, to study the pyrazine $S_0 \to S_2 \leftrightarrow S_1$ excitation and IC dynamics. Using this as a guide, we computed  control and dynamics in the corresponding energy range ($E_L$ = 4.68 eV, $E_H$ = 4.96 eV), using $T_1$ = 150 fs, $T_2$ = 250 fs, $N_A$ = 128, and all $\alpha_a$ = 21.0 fs. The resulting $S_2$ populations, together with resulting control fields in time domain, are shown in Fig. \ref{Figure-P_c_t_128_bins} where the subscript c denotes ``controlled". Corresponding control fields in the frequency domain are shown in Figs. \ref{Figure-varepsilon_min_128_bins} and \ref{Figure-varepsilon_max_128_bins}.

The behavior of the controlled $P_{S_2} (t)$  (Fig. \ref{Figure-P_c_t_128_bins}), differs in magnitude in the regions when the pulse is acting, and after the pulse is over. To understand this difference, note that to obtain  the  controlled fields in Figs. \ref{Figure-varepsilon_min_128_bins} and \ref{Figure-varepsilon_max_128_bins} using a set of Gaussians requires that some components of $\mathbf{d}$ vector  be large. After the pulse is over, these components are ``balanced'' by one another in the \textit{infinite-time} Fourier transform, to give  the small desired population value $P_0$ at $t = T_1$ or $t = T_2$ and to yield the required controlled dynamics. However, while the pulse is acting, these components are ``unbalanced''  giving large transient $\varepsilon_p(\omega,t)$ values.  For similar reasons
 the controlled pulses, being a linear combinations of single Gaussians, are effectively longer than the single Gaussian pulse (see Fig. \ref{Figure-P_c_t_128_bins}, lower panel).

 To examine  the complex structure of the control pulses at  Figs. \ref{Figure-varepsilon_min_128_bins} and \ref{Figure-varepsilon_max_128_bins}, we apply several approaches to simplify the field  while monitoring the control achieved. First, we attempted a local averaging of the controlled field, where the total field in $N_A$ bins is arithmetically averaged (amplitude and phase separately) using a smaller number $N_S$ of larger bins ($N_A$ being an integer multiple of $N_S$, for example, for $N_A = 64$, $N_S$ = 32, 16, 8, 4, 2). By doing so, the resulting averaged field, however,  showed virtually no control. Second, this averaged step-like field was expanded with $N_S$ Gaussians and the resulting smoothed field  used for the propagation.  Again, this case led to nearly complete loss of control.

\begin{figure}[htp]
\begin{center}
\includegraphics[height = 14cm, width = 12cm]{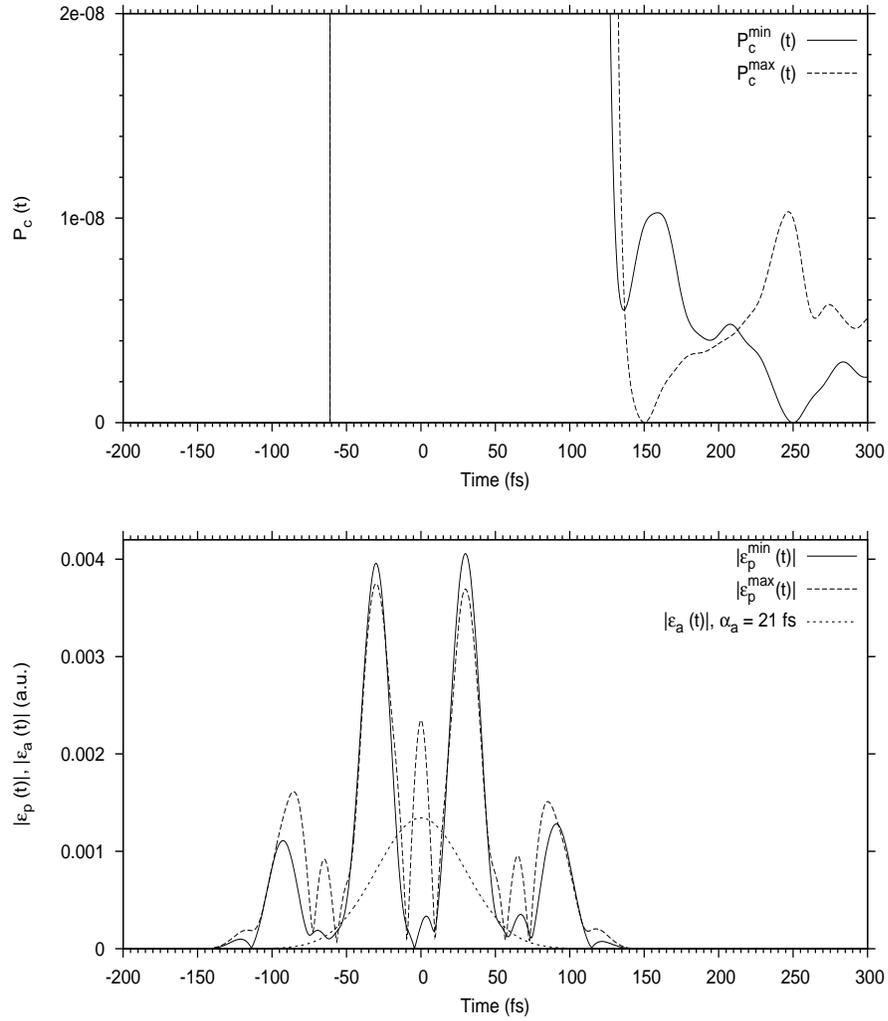}
\caption{Upper panel: Two controlled $S_2$ populations, $P_{S_2}(t)$, denoted $P_c (t)$, which either minimize and maximize $\lambda^{R, \mathbf{A}}$, i.e., the $S_2$ population ratio at times $T_2$ = 250 fs and $T_1$ = 150 fs.   The $P^{max}_c$ curve has been multiplied by $8.3 \times 10^{-5}$ in order to fit on this figure.   Lower panel: Time envelopes of two corresponding controlled laser pulses, $|\varepsilon_p (t)|$, together with the time envelope of the single (uncontrolled) Gaussian laser pulse, $|\varepsilon_a (t)|$.} \label{Figure-P_c_t_128_bins}
\end{center}
\end{figure}

\begin{figure}[htp]
\begin{center}
\includegraphics[height = 7cm, width = 10cm]{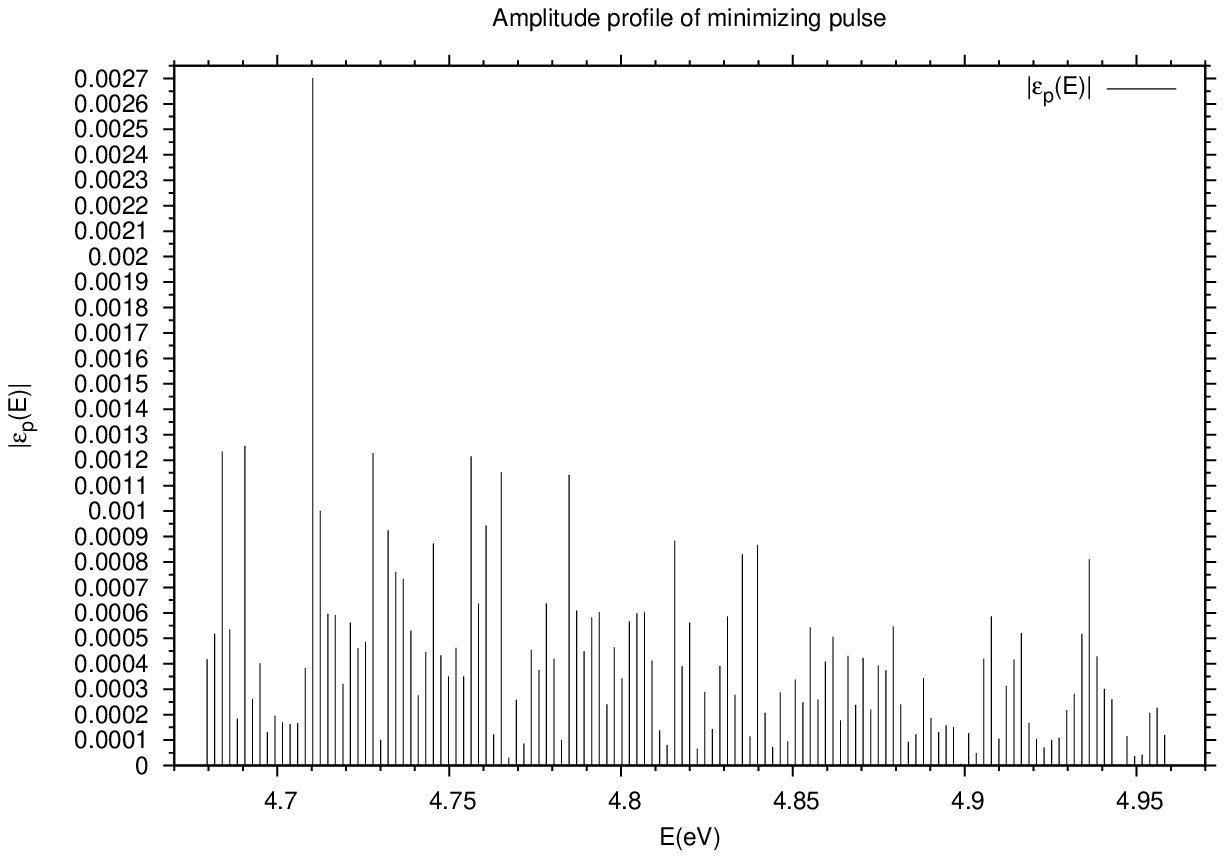}
\includegraphics[height = 7cm, width = 10cm]{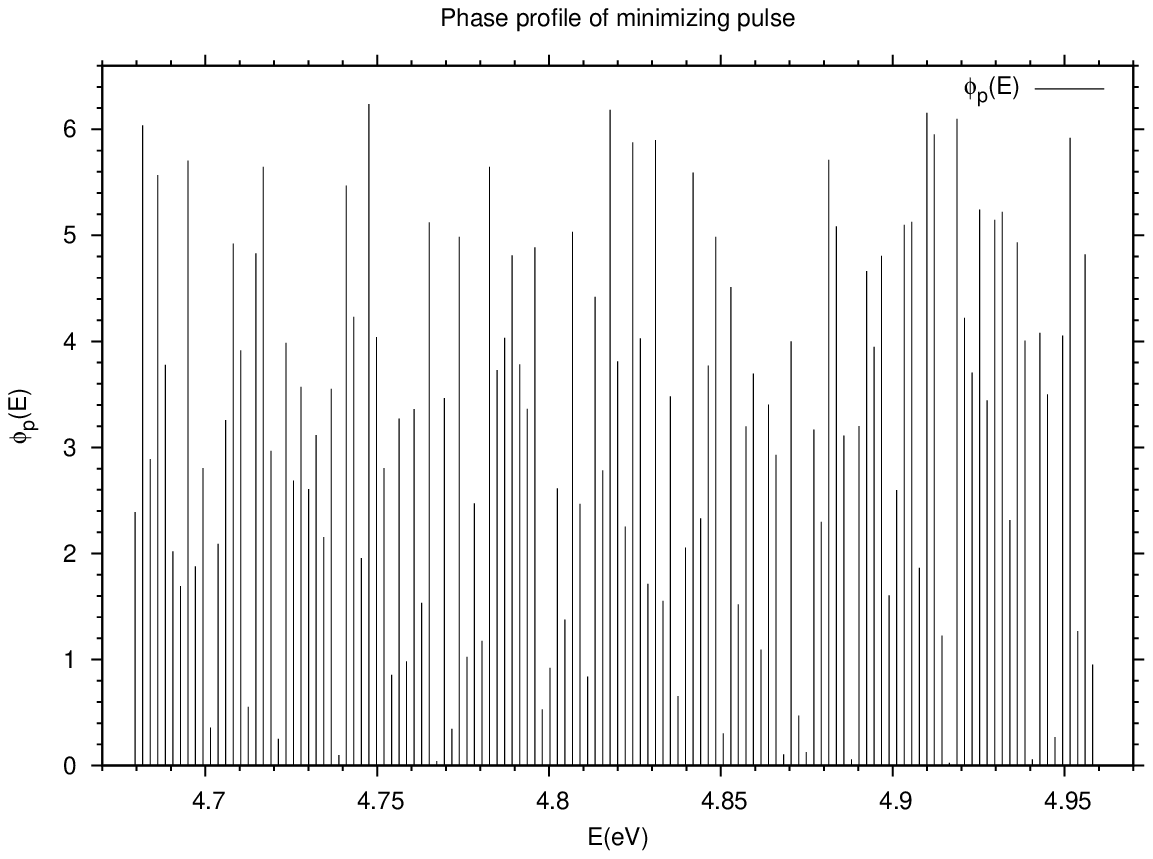}
\caption{Amplitude and phase of $\underline{\varepsilon}^\mathbf{A}_p$ eigenvector, which minimizes the $P_{S_2} (T_2) / P_{S_2} (T_1)$ ratio. $\lambda^{R, \mathbf{A}}_{\min}$ = 8.28$\times$10$^{-5}$.} \label{Figure-varepsilon_min_128_bins}
\end{center}
\end{figure}

\begin{figure}[htp]
\begin{center}
\includegraphics[height = 7cm, width = 10cm]{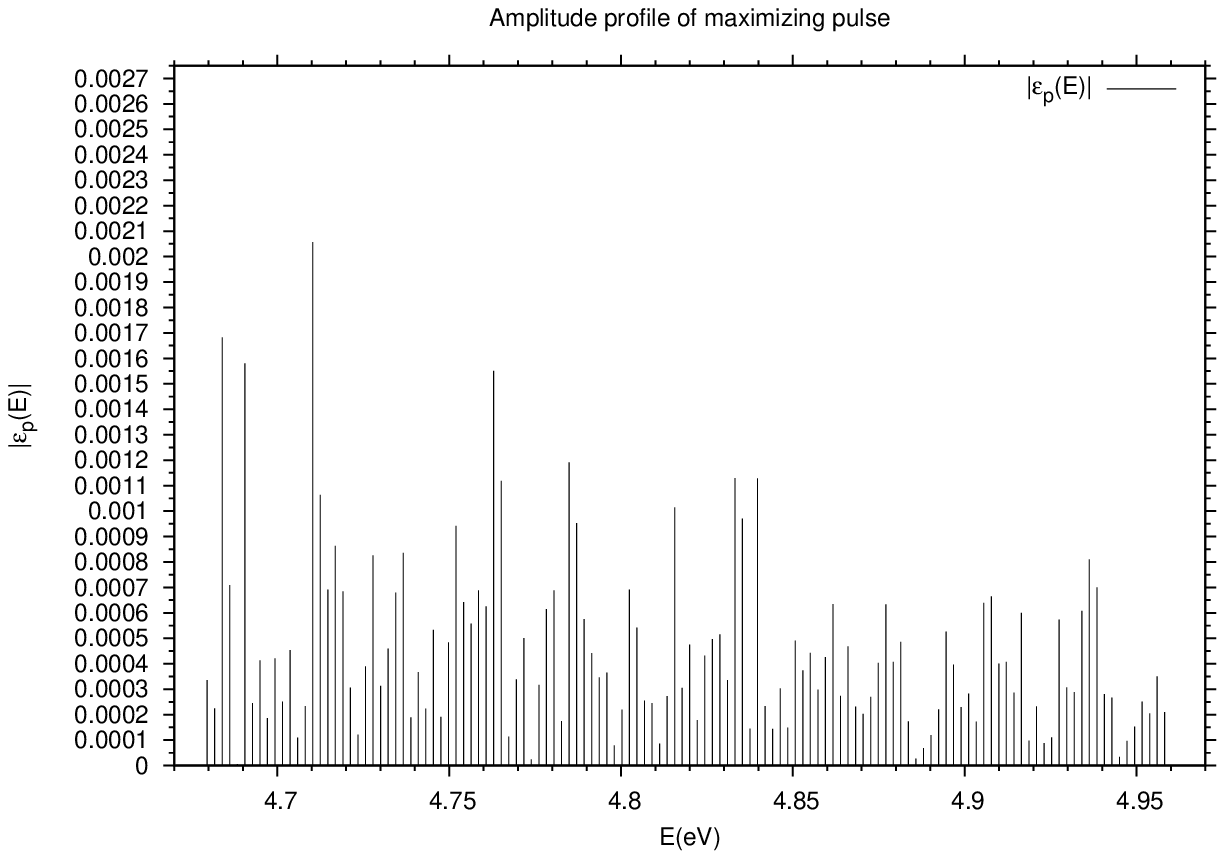}
\includegraphics[height = 7cm, width = 10cm]{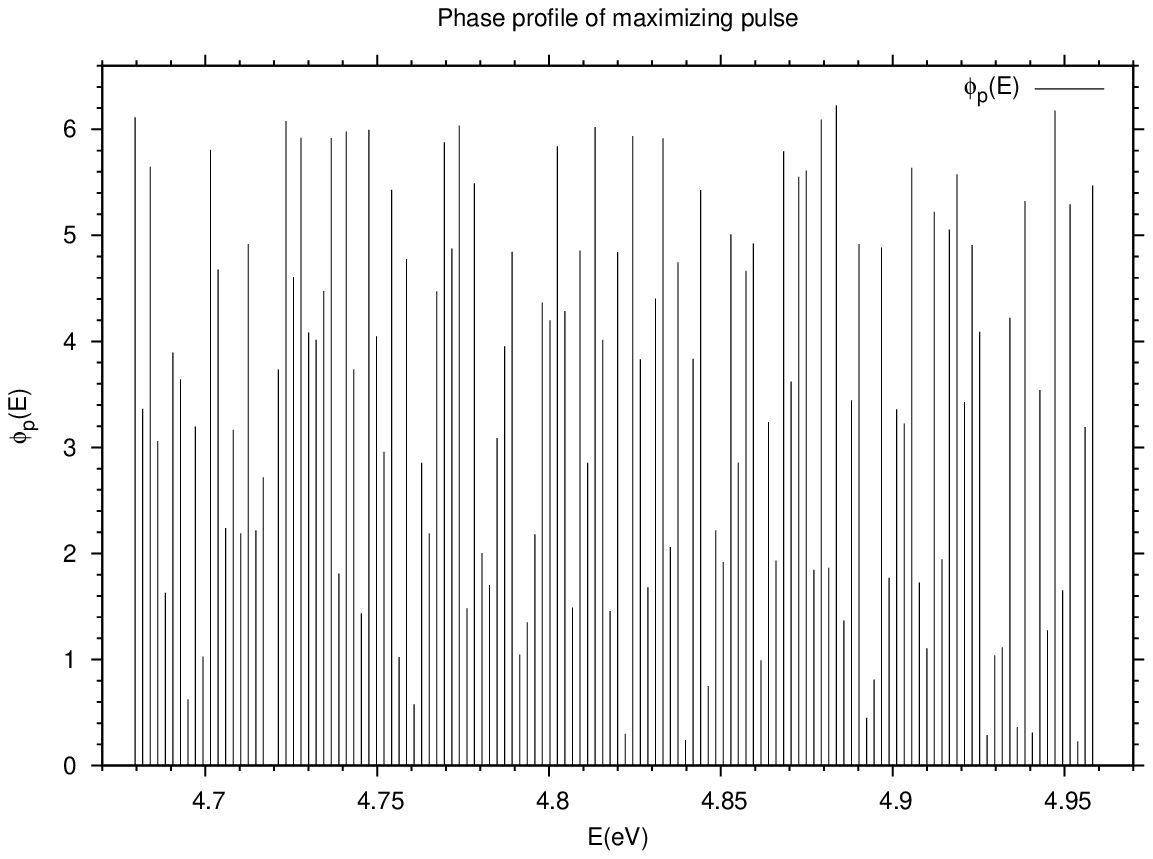}
\caption{Amplitude and phase of $\underline{\varepsilon}^\mathbf{A}_p$ eigenvector, which maximizes the $P_{S_2} (T_2) / P_{S_2} (T_1)$ ratio. $\lambda^{R, \mathbf{A}}_{\max}$ = 8.38$\times$10$^{+3}$.} \label{Figure-varepsilon_max_128_bins}
\end{center}
\end{figure}

\begin{figure}[t]
\begin{center}
\includegraphics[height = 9cm, width = 12cm]{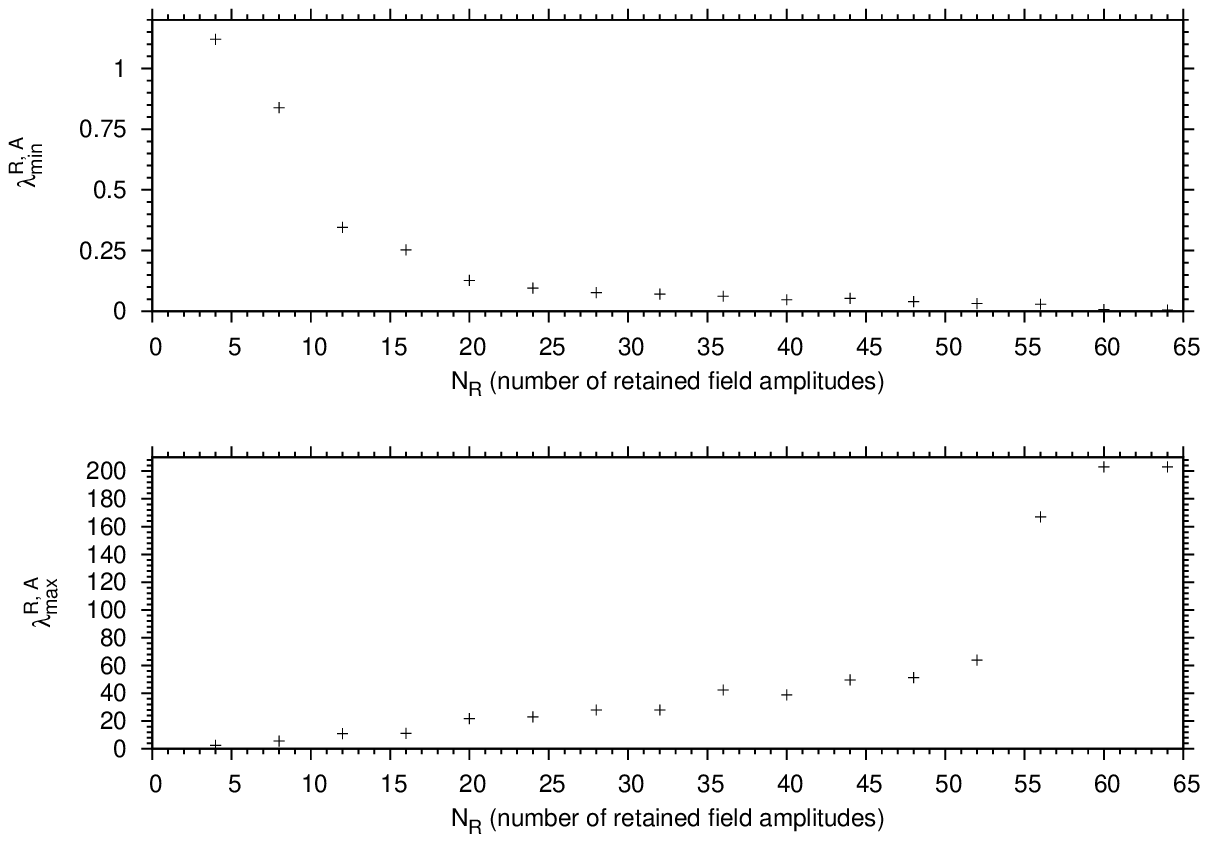}
\caption{Upper panel: Dependence of $\lambda^{R, \mathbf{A}}_{\min}$ on the number of retained amplitudes $N_R$. Lower panel: The same, but for $\lambda^{R, \mathbf{A}}_{\max}$. Total number of amplitudes $N_A$ = 64.} \label{Figure-N_R_64_bins}
\end{center}
\end{figure}

 An alternative simplifying approach was, however, successful.  Specifically,  we retained only the $N_R$ largest field amplitudes out of the total $N_A$ (with all the smaller ampitudes set to zero), keeping the phase profile intact, and monitoring the changes in control ratios. Sample results for $N_A$ = 64 are shown in Fig. 7.   A total $N_A$ of  64 is used here  (results with $N_A$ = 128  are qualitatively the same). It is clear from Fig. \ref{Figure-N_R_64_bins}, that this approach, retaining only the largest amplitudes, works better than the previous two since it tends to partially maintain important dynamical information. Generally, $\lambda^{R, \mathbf{A}}_{\min}$ is more robust with respect to this amplitude truncation than is $\lambda^{R, \mathbf{A}}_{\max}$. Additionally, we found that the extent of control achieved using only $N_R$ amplitudes out of $N_A$, is similar in magnitude to control extents without truncation, but using this $N_R$ as the original  $N_A$. That is,
  the same number of degrees of freedom in both cases provides similar extents of control.

Theoretically, maximum and minimum control limits via this approach can be reached using all coarse-grained $| \overline{\alpha} \rangle$ states accessible to the laser, \textit{i.e.}, those belonging to the interval of interest $[E_L, E_H]$. For the  case presented in Figs. \ref{Figure-P_c_t_128_bins}, \ref{Figure-varepsilon_min_128_bins} and \ref{Figure-varepsilon_max_128_bins}, the number of  $| \overline{\alpha} \rangle$ states, using our pyrazine description, is 11885. However, as mentioned in the Appendix, the optimization problem for $| \overline{\alpha} \rangle$ states in Eq. (\ref{RelativeControlProblem-alpha}) is numerically stable only up to dimensionality 150--180, and the control range $\lambda^{R, \alpha}_{\max} - \lambda^{R, \alpha}_{\min}$ continues to increase when the  dimensionality increases from 128 to 180, reaching $\sim 10^5$. We anticipate a theoretical control range limit to be  $\sim 10^{9}$--$10^{10}$, which, however, is not achieved due to the numerical limitations discussed in Appendix (see below).

\section{Summary and Conclusions} \label{Summary}

Coherent control of internal conversion (IC) between the first and second singlet excited electronic states of pyrazine ($S_1$ and $S_2$) is examined, using two different control objectives. The control is performed by means of shaping the laser, which excites the system from the ground electronic state $S_0$ to the second excited electronic state $S_2$. Resonance energy broadening and resonance overlap are shown to be responsible for phase control efficiency, and a correlation between resonance overlap and controllability  is established. A huge range of control was obtained for the relative population of $S_2$ at long times as compared to times just after the pulse is over.  Different ways to simplify the controlled fields are described, and the behavior of the control  as a consequence of these simplifications is investigated. Specifically, we have found that retaining the largest field amplitudes is the best approach to field simplification.

\section{Acknowledgements}

This work was supported by the Natural Sciences and Engineering Research Council of Canada (NSERC).  This manuscript summarizes one of the last joint efforts of the Brumer and Shapiro research groups.  The topic, overlapping resonance effects, was beloved by Moshe.  P.B. is grateful for the opportunity to have interacted with such an outstanding scientist for over 40 years, publishing over 120 joint papers, and two books.

\newpage

\noindent  \textbf{Appendix:  Energy Partitioning for Control Associated with Overlapping Resonances} \label{Theory-Control-A-Partitioning}

As noted in Sect. \ref{Theory-Control-Overlapping-Resonances}, numerical instabilities necessitated that we introduce a further
partitioning of the energy axis. Specifically,
we partitioned  the energy axis   into a limited number of $N_A$ bins in Eq. (\ref{kappa_Psi_p_Overlap}):

\begin{eqnarray}
\langle \kappa | \Psi_p (t) \rangle & \approx & \sum_A \sum_{\alpha \in I_A} \varepsilon_p (\omega_{\alpha,g},t) M^{\varepsilon, \alpha}_{\kappa,\alpha}(t) \nonumber \\
& \approx & \sum_A \varepsilon_p (\omega_{A,g},t) \left[ \sum_{\alpha \in I_A} M^{\varepsilon, \alpha}_{\kappa,\alpha}(t) \right] \equiv \sum_A \varepsilon_p (\omega_{A,g},t) M^{\varepsilon, \mathbf{A}}_{\kappa,A}(t), \label{kappa_Psi_p_BinnedOverlap}
\end{eqnarray}
where $I_A$ is a bin number A, which has the center energy $E_A$, $\omega_{A,g} \equiv (E_A - E_g)/\hbar$, and
\begin{equation}
M^{\varepsilon, \mathbf{A}}_{\kappa,A}(t) \equiv \sum_{\alpha \in I_A} M^{\varepsilon, \alpha}_{\kappa,\alpha}(t)
\label{M_kappa_A_def}
\end{equation}
is the collective material system matrix element, corresponding to  bin $I_A$. Here and below, the bold superscript $\mathbf{A}$  denotes that all the corresponding quantities are written for $| A \rangle$ states, which are defined below.

Using Eq. (\ref{M_varepsilon_kappa_alpha}), $M^{\varepsilon, \mathbf{A}}_{\kappa,A}(t)$ can be written as:
\begin{equation}
M^{\varepsilon, \mathbf{A}}_{\kappa,A}(t) = \sum_{\alpha \in I_A} \langle \kappa | \overline{\alpha} \rangle \, \tau_\alpha(t) \, \frac{i}{\hbar} \sum_{\kappa'} \langle \overline{\alpha} | \kappa' \rangle \langle \kappa' | \mu | g \rangle \equiv \langle \kappa | \left[ \sum_{\alpha \in I_A} \tau_\alpha (t) | \overline{\alpha} \rangle \langle \overline{\alpha} | \right] \left[ \sum_{\kappa'} \frac{i}{\hbar} \langle \kappa' | \mu | g \rangle | \kappa' \rangle \right]. \label{M_varepsilon_propagator}
\end{equation}
The middle expression in square brackets, unlike Eq. (\ref{M_varepsilon_kappa_alpha}), is not the single $| \overline{\alpha} \rangle$ state propagator, but the localized coarse-grained propagator, with the sum only over $| \overline{\alpha} \rangle$ states belonging to the  bin $I_A$.

One can introduce the ``binned" states $| A \rangle$, such that the corresponding projector onto the state $| A \rangle$ is
$$ | A \rangle \langle A | = (1/N_{I_A}) \sum_{\alpha \in I_A} | \overline{\alpha} \rangle \langle \overline{\alpha} |,~~{\rm hence}~~ \sqrt{N_{I_A}} | A \rangle \langle A | \sqrt{N_{I_A}} = \sum_{\alpha \in I_A} | \overline{\alpha} \rangle \langle \overline{\alpha} |, $$
where $N_{I_A}$ is the number of $| \overline{\alpha} \rangle$ states that are inside bin $I_A$.  With  the notation $| \overline{A} \rangle \equiv \sqrt{N_{I_A}} | A \rangle$ we have
\begin{equation}
| \overline{A} \rangle \langle \overline{A} | = \sum_{\alpha \in I_A} | \overline{\alpha} \rangle \langle \overline{\alpha} |. \label{A_bar_projector_definition}
\end{equation}
Using Eq. (\ref{A_bar_projector_definition}), the propagator in Eq. (\ref{M_varepsilon_propagator}) can be approximately rewritten as:
\begin{equation}
\sum_{\alpha \in I_A} \tau_\alpha (t) | \overline{\alpha} \rangle \langle \overline{\alpha} | \approx | A \rangle \langle A | \sum_{\alpha \in I_A} \tau_\alpha (t) \equiv \tau^\mathbf{A}_A (t) | \overline{A} \rangle \langle \overline{A} |, \label{Propagator_binned_approximation}
\end{equation}
where $\tau^\mathbf{A}_A (t) \equiv (1/N_{I_A}) \sum_{\alpha \in I_A} \tau_\alpha (t)$. The accuracy of the approximation made in Eq. (\ref{Propagator_binned_approximation}) rapidly increases with  decreasing  bin size. Using Eq. (\ref{Propagator_binned_approximation}), $M^{\varepsilon,\mathbf{A}}_{\kappa,A} (t)$ in Eq. (\ref{M_varepsilon_propagator}) can be rewritten as:
\begin{equation}
M^{\varepsilon,\mathbf{A}}_{\kappa,A}(t) \approx \langle \kappa | \overline{A} \rangle \tau^\mathbf{A}_A (t) \frac{i}{\hbar} \sum_{\kappa'} \langle \overline{A} | \kappa' \rangle \langle \kappa' | \mu | g \rangle.
\end{equation}
Defining
\begin{eqnarray}
\mu^{\mathbf{A}}_A & \equiv & \frac{i}{\hbar} \sum_{\kappa'} \langle \overline{A} | \kappa' \rangle \langle \kappa' | \mu | g \rangle = \frac{i}{\hbar} \langle \overline{A} | \mu | g \rangle, \label{mu_A_def} \qquad R^{\mathbf{A}}_{A,\kappa} \equiv \langle \overline{A} | \kappa \rangle
\end{eqnarray}
gives
\begin{equation}
\mathbf{M}^{\varepsilon, \mathbf{A}} (t) = \mathbf{R}^{\mathbf{A} \dagger} \, \underline{\underline{\tau}}^\mathbf{A} (t) \, \underline{\underline{\mu}}^\mathbf{A}, \label{M_A_varepsilon_t_definition} \qquad \mathbf{K}^{\varepsilon, \mathbf{A}} (t) = \mathbf{M}^{\varepsilon, \mathbf{A} \dagger} (t) \mathbf{M}^{\varepsilon, \mathbf{A}} (t) = \underline{\underline{\mu}}^{\mathbf{A} \dagger} \underline{\underline{\tau}}^{\mathbf{A} \dagger} (t) \mathbf{R}^\mathbf{A} \mathbf{R}^{\mathbf{A} \dagger} \, \underline{\underline{\tau}}^\mathbf{A} (t) \, \underline{\underline{\mu}}^\mathbf{A}, \label{K_A_varepsilon_t_definition}
\end{equation}
where $\underline{\underline{\tau}}^\mathbf{A} (t)$ is a square diagonal matrix composed of $\tau^\mathbf{A}_A (t)$ values, and $\underline{\underline{\mu}}^\mathbf{A}$ is a square diagonal matrix composed of $(i/\hbar) \langle \overline{A} | \mu | g \rangle$ values. This gives the $P_{S_2} (t)$ population in terms of binned values as
\begin{eqnarray}
  P_{S_2} (t)& = & \underline{\varepsilon}^{\mathbf{A} \dagger} (t) \underline{\underline{\mu}}^{\mathbf{A} \dagger} \underline{\underline{\tau}}^{\mathbf{A} \dagger} (t) \mathbf{R}^\mathbf{A} \mathbf{R}^{\mathbf{A} \dagger} \, \underline{\underline{\tau}}^\mathbf{A} (t) \, \underline{\underline{\mu}}^\mathbf{A} \underline{\varepsilon}^\mathbf{A} (t)
   \\ & = &\sum_{A'} | \varepsilon_p (\omega_{A',g},t) |^2 K^{\varepsilon, \mathbf{A}}_{A',A'} (t) +  \sum_{A' \ne A''} \! \! \! \varepsilon_p^* (\omega_{A',g},t) \varepsilon_p (\omega_{A'',g},t) K^{\varepsilon, \mathbf{A}}_{A',A''} (t), \label{P_S_2_t_A_matrix_expansion}
\end{eqnarray}
where $\underline{\varepsilon}^\mathbf{A} (t)$ is a vector composed of $\varepsilon_p (\omega_{A,g},t)$ components.

Since $\underline{\underline{\mu}}^\mathbf{A}$ and $\underline{\underline{\tau}}^\mathbf{A} (t)$ are diagonal, the only possible source of nondiagonality in Eq. (\ref{K_A_varepsilon_t_definition}) for $\mathbf{K}^{\varepsilon, \mathbf{A}} (t)$ and Eq. (\ref{P_S_2_t_A_matrix_expansion}) is $\mathbf{Q}^\mathbf{A} = \mathbf{R}^\mathbf{A} \mathbf{R}^{\mathbf{A} \dagger}$, composed of $Q^\mathbf{A}_{A',A''} = \langle \overline{A}' | Q | \overline{A}'' \rangle$ values. Thus, the possibility of phase control by means of phases $\phi_A (t)$ of complex $\varepsilon_p (\omega_{A,g},t) = |\varepsilon_p (\omega_{A,g},t)| \exp(i \phi_A (t))$ depends solely on its properties. As in the previous case of $| \gamma \rangle$ and $| \overline{\alpha} \rangle$, all the $P_{S_2} (t)$ phase control considerations remain the same, except that $| \gamma \rangle$ or $| \overline{\alpha} \rangle$ states are replaced by $|\overline{A} \rangle$ states. Namely, phase control is provided by resonance energy broadening and resonance overlap.  The resonance overlap effect, providing the non-block-diagional structure of $\mathbf{Q}^\mathbf{A}$ and $\mathbf{K}^{\varepsilon, \mathbf{A}} (t)$ as a consequence, enhances the effect of resonance broadening.

Using Eq. (\ref{P_S_2_t_A_matrix_expansion}), the eigenproblem in Eq. (\ref{RelativeControlProblem-alpha}) is reformulated as
\begin{eqnarray}
\mathbf{R}^{\varepsilon, \mathbf{A}} (T_2, T_1) \underline{\varepsilon}^\mathbf{A} & = & \lambda^{R, \mathbf{A}} \underline{\varepsilon}^\mathbf{A}, \label{RelativeControlProblem-A-App} \\
\mathbf{R}^{\varepsilon, \mathbf{A}} (T_2, T_1) & \equiv & [\mathbf{K}^{\varepsilon, \mathbf{A}} (T_1)]^{-1} \mathbf{K}^{\varepsilon, \mathbf{A}} (T_2). \label{RelativeControlMatrix-A-App}
\end{eqnarray}
Its dimensionality reduced from $N_\alpha$ to $N_A$, allowing an accurate numerical solution for $N_A$ values up to 150-180.





\begin{thebibliography}{0.5cm}

\bibitem{address}  Current address:  Advanced Chemistry Development, Inc., 8 King Street East, Toronto, Ontario, Canada M5C 1B5

\bibitem{Rice-Zhao-2000}
S. A. Rice and M. Zhao, \textit{Optical Control of Molecular Dynamics} (Wiley, New York, 2000).

\bibitem{Shapiro-Brumer-2003}
M. Shapiro and P. Brumer, \textit{Principles of the Quantum Control of Molecular Processes} (Wiley, New York, 2003);
M. Shapiro and P. Brumer, \textit{Quantum Control of Molecular Processes} (Wiley-VCH, Weinheim, 2012).

\bibitem{Frishman-Shapiro-2001}
E. Frishman and M. Shapiro, Phys. Rev. Lett. \textbf{87}, 253001 (2001).

\bibitem{Collinear-OCS}
P. S. Christopher, M. Shapiro and P. Brumer, J. Chem. Phys. \textbf{126}, 124307 (2007).

\bibitem{Christopher-Pyrazine-1}
P. S. Christopher, M. Shapiro and P. Brumer, J. Chem. Phys. \textbf{123}, 064313 (2005).

\bibitem{Christopher-Pyrazine-2}
P. S. Christopher, M. Shapiro and P. Brumer, J. Chem. Phys. \textbf{124}, 184107 (2006).

\bibitem{Christopher-Pyrazine-3}
P. S. Christopher, M. Shapiro and P. Brumer, J. Chem. Phys. \textbf{125}, 124310 (2006).

\bibitem{Borrelli-Peluso-2003}
R. Borrelli and A. Peluso, J. Chem. Phys. \textbf{119}, 8437 (2003).

\bibitem{Levine-1969}
R. D. Levine, \textit{Quantum Mechanics of Molecular Rate Processes} (Clarendon Press, Oxford, 1969).

\bibitem{Shapiro-1972}
M. Shapiro, J. Chem. Phys. \textbf{56}, 2582 (1972).

\bibitem{Raab-1999}
A. Raab, G. A. Worth, H.-D. Meyer and L. S. Cederbaum, J. Chem. Phys. \textbf{110}, 936 (1999).

\bibitem{IBr-model}
T. Grinev, M. Shapiro and P. Brumer, J. Chem. Phys. \textbf{138}, 044306 (2013).

\bibitem{Shapiro-1998}
M. Shapiro, J. Phys. Chem. A \textbf{102}, \textit{47}, 9570 (1998).

\bibitem{ChemPhysLett-2009}
R. He, C. Zhu, C.-H. Chin and S. H. Lin, Chem. Phys. Lett. \textbf{476}, 19 (2009).

\bibitem{Preisig}
J. C. Preisig, SIAM J. Control  Optim. \textbf{34}, 1135 (1996).

\bibitem{Kosloff-1989}
R. Kosloff, S. A. Rice, P. Gaspard, S. Tersigni and D. J. Tannor, Chem. Phys. \textbf{139}, 201 (1989).

\bibitem{Shapiro-1993}
M. Shapiro, J. Phys. Chem. \textbf{97}, \textit{29}, 7396 (1993).

\bibitem{Shapiro-Femtosecond}
M. Shapiro, in \textit{Femtosecond Chemistry}; J. Manz, L. Woste, Eds. (VCH, Weinheim, 1995); p. 321.

\bibitem{Wolfram-Erf}
E. W. Weisstein, Erf(z) error function. From MathWorld--A Wolfram Web Resource. \\
http://mathworld.wolfram.com/Erf.html

\bibitem{Abramowitz-Stegun}
M. Abramowitz and I. A. Stegun, \textit{Handbook of Mathematical Functions} (Dover, New York, 1965); Eqs. 7.1.3 and 7.1.8.

\bibitem{Batista-2006}
X. Chen and V. S. Batista, J. Chem. Phys. \textbf{125}, 124313 (2006).

\bibitem{Pyrazine-Ioannis-2}
I. Thanopulos, X. Li, P. Brumer and M. Shapiro,  J. Chem. Phys. \textbf{137}, 064111 2012.

\bibitem{Pyrazine-Ioannis-1}
I. Thanopulos, P. Brumer and M. Shapiro, J. Chem. Phys. \textbf{133}, 154111 (2010).

\end{thebibliography}
\end{document}